\documentclass[lettersize,journal]{IEEEtran}
\usepackage{cite}
\usepackage[dvipsnames, table]{xcolor}
\usepackage{colortbl}

\makeatletter
\def\@IEEEstripcolor#1{\color{#1}}
\makeatother

\usepackage{amsmath,amssymb,amsfonts}
\usepackage{bm}
\usepackage{optidef}
\usepackage{mathptmx} % Times font for math

\DeclareMathOperator*{\argmin}{arg\,min}

\usepackage{algorithm}
\usepackage{algpseudocode}

\usepackage{booktabs}
\usepackage{makecell}
\usepackage{multirow}
\usepackage{tabularx}
\usepackage{array}
\usepackage{hhline}
\usepackage{arydshln}

\usepackage{adjustbox}

\usepackage{graphicx}
\usepackage{wrapfig}
\usepackage{adjustbox}
\usepackage{caption}
\usepackage{subcaption}

\usepackage{textcomp}
\usepackage{xurl}
\usepackage{indentfirst}
\usepackage{lipsum}
\usepackage{soul}        % for \hl
\usepackage{soulpos}     % for \ulpos, etc.
\usepackage[normalem]{ulem} % SAFE: keeps italics for \emph

\newif\ifshownormalchanges
\shownormalchangesfalse   % or \shownormalchangesfalse or set to true for track changes (revision 1)

\newif\ifshowsuperchanges
\showsuperchangesfalse   % or \showsuperchangesfalse set to true for track changes (revision 2)

\newcommand{\added}[1]{%
  \ifshownormalchanges{\textcolor{blue}{#1}}\else#1\fi}

\newcommand{\deleted}[1]{%
  \ifshownormalchanges{\textcolor{red}{\sout{#1}}}\fi}

\newcommand{\superadded}[1]{%
  \ifshowsuperchanges{\textcolor{blue}{#1}}\else#1\fi}

\newcommand{\superdeleted}[1]{%
  \ifshowsuperchanges{\textcolor{red}{\sout{#1}}}\fi}

\usepackage[hidelinks]{hyperref}

\begin{document}

\title{\deleted{Efficient Traffic Classification using HW-NAS: Advanced Analysis and Optimization for Cybersecurity on Resource-Constrained Devices}}
\title{\added{Hardware-Aware Neural Architecture Search for Encrypted Traffic Classification on Resource-Constrained Devices}}
%\author{IEEE Publication Technology,~\IEEEmembership{Staff,~IEEE,}
        % <-this % stops a space
\author{
Adel Chehade,~\IEEEmembership{Graduate Student Member,~IEEE,}
Edoardo Ragusa,~\IEEEmembership{Member,~IEEE,}
Paolo Gastaldo,~\IEEEmembership{Member,~IEEE,}
and Rodolfo Zunino
\thanks{The authors are with the Department of Naval, Electrical, Electronic and Telecommunications Engineering (DITEN), University of Genoa, 16126 Genoa, Italy (e-mail: adel.chehade@edu.unige.it; edoardo.ragusa@unige.it; paolo.gastaldo@unige.it; rodolfo.zunino@unige.it).}

\thanks{This work was partially supported by project SERICS (PE00000014) under the MUR National Recovery and Resilience Plan funded by the European Union - NextGenerationEU.}

\thanks{© 2026 The Authors. Licensed under the Creative Commons Attribution 4.0 License (CC BY 4.0). The final version of this article appears in \emph{IEEE Transactions on Network and Service Management} and is available at https://doi.org/10.1109/TNSM.2026.3666676.}
}


\maketitle

\begin{abstract}

This paper presents a hardware-efficient deep neural network (DNN), optimized through hardware-aware neural architecture search (HW-NAS); the DNN supports the classification of session-level encrypted traffic on resource-constrained Internet of Things (IoT) and edge devices. Thanks to HW-NAS, a 1D convolutional neural network (CNN) is tailored on the ISCX VPN-nonVPN dataset to meet strict memory and computational limits while achieving robust performance. The optimized model attains an accuracy of \deleted{96.59\%}\added{96.60\%} with just 88.26K parameters, 10.08M \added{floating-point operations (FLOPs)}, and a maximum tensor size of 20.12K. Compared to state-of-the-art (SOTA) models, it achieves reductions of up to 444-fold, 312-fold, and 15-fold in these metrics, respectively, significantly minimizing memory footprint and runtime requirements. \deleted{The model also demonstrates versatility in classification tasks, achieving accuracies of up to 99.64\% in VPN differentiation, VPN-type classification, broader traffic categories, and application identification}\added{The model also demonstrates versatility, achieving up to 99.86\% across multiple VPN and traffic classification (TC) tasks}\added{; it further generalizes to external benchmarks with up to 99.98\% accuracy on USTC-TFC and QUIC NetFlow.}
In addition, an in-depth approach to header-level preprocessing strategies confirms that the optimized model can provide notable performance across a wide range of configurations, even in scenarios with stricter privacy considerations.
Likewise, a reduction in the length of sessions of up to 75\% yields significant improvements in efficiency, while maintaining high accuracy with only a negligible drop of 1-2\%. However, the importance of careful preprocessing and session length selection in the classification of raw traffic data is still present, as improper settings or aggressive reductions can bring about a 7\% reduction in overall accuracy.
\added{The quantized architecture was deployed on STM32 microcontrollers and evaluated across input sizes; results confirm that the efficiency gains from shorter sessions translate to practical, low-latency embedded inference.}
\deleted{Those results highlight the method's effectiveness in enforcing cybersecurity for IoT networks, by providing scalable, efficient solutions for the real-time  analysis of encrypted traffic within strict hardware limitations.}\added{These findings demonstrate the method’s practicality for encrypted traffic analysis in constrained IoT networks.}

\end{abstract}


%\titlepgskip=-15pt

\maketitle

\section{Introduction}
\label{sec:introduction}

\IEEEPARstart{T}{he} proliferation of Internet of Things (IoT) technologies introduces security challenges that traditional methods often cannot handle effectively \cite{iot_6g_martalo2024cross}. Resource-constrained devices generate huge amounts of data; relying on centralized servers to process that data may lead to transfer delays, increased network load, and additional power consumption \cite{gallo2021fenxi}. Ideally, data flow monitoring should be carried out on edge devices to limit overhead in network management \cite{pacheco2018towards}. This paper proposes a design strategy to enable efficient traffic classification (TC) on constrained devices \added{at the network edge}.

The growth of encrypted Internet traffic, which now accounts for up to 96\% of data \cite{survey_DONG2024128444}, worsens this scenario. Widespread encryption complicates network security and traffic analysis \cite{seydali2023cbs, lotfollahi2020deep}; since modern protocols conceal packet contents, traditional methods such as Deep Packet Inspection (DPI) and port-based classification are mostly ineffective \cite{pacheco2018towards}.

Efficient real-time traffic analysis is essential in today’s cybersecurity ecosystem, especially for devices with limited resources like IoT and edge platforms \cite{gallo2021fenxi}. These devices require fast, adaptive processing to manage encrypted traffic patterns and support real-time decision-making. Use cases include \added{edge monitoring in} high-bandwidth 5G applications, such as video streaming and gaming, which need smooth performance \cite{adil20245g}, and sensitive areas like finance and healthcare, where secure data exchange is critical \cite{survey_DONG2024128444}.

Deep neural networks (DNNs) are increasingly used for traffic classification (TC) due to their high accuracy and ability to automatically extract features \cite{nas_etc_malekghaini2023automl4etc, seydali2023cbs}. However, their high computational needs limit their use in devices with limited resources \cite{tabanelli2023dnn}. This work explores ways to optimize DNN designs specifically for these restricted settings.

Beyond the DNN architecture, optimizing classification performance requires careful design choices. Network traffic can be classified at different levels, namely packets, flows, and sessions \cite{li2024l2}. A packet is the smallest unit of data, while a flow includes packets sent from a source to a destination. A session extends this concept by including bidirectional communication between two endpoints, capturing the full packet exchange over time. This paper focuses on session-level classification, as it provides a broader view of network behavior, particularly for identifying encrypted traffic patterns \cite{wang2017end, centime_maonan2021centime, attention_spatio_hu2023network, zou2024novel_residual}.

Another important aspect is the processing of packets; this process aims to ensure that only relevant information feeds the neural network. Several approaches in the literature \cite{xu2024cascaded,yao2019capsule,bu2020encrypted_parallel_packet,yao2019identification} considered both accuracy and computational costs.

Finally, one should consider that the varying levels of complexity in the overall design problem can make certain strategies more feasible than others. This paper analyzes the interaction between those factors from an empirical viewpoint and provides the reader with practical design guidelines.

The paper presents a solution that relies on hardware-aware neural architecture search (HW-NAS) for network TC. \superdeleted{The use of HW-NAS to optimize DNNs for processing raw encrypted data introduces a novel approach to cybersecurity and network traffic analysis.}HW-NAS has been applied in various deep learning (DL) fields \cite{1000papersNAS}, \cite{chitty2023neural_ieee-access}; \superadded{prior studies have shown that hardware objectives are most effectively handled during architecture optimization, rather than through post-hoc compression or manual tuning \cite{tan2019mnasnet,cai2018proxylessnas}. In this work, HW-NAS is employed as a constraint-driven design methodology, in which parameter count, floating-point operations (FLOPs), and peak intermediate tensor size are explicitly bounded during the search. These constraints reflect practical deployment bottlenecks in model storage, computational cost, and runtime memory usage.}

The application considered in this paper deals with a significant, uncovered aspect in the literature, that is, the design of optimized neural models that satisfy strict hardware constraints and yet maintain satisfactory classification performances across diverse network environments\superdeleted{. HW-NAS can account for limitations in memory, computational power, and energy consumption}, thus making real-world deployment on IoT (edge) platforms feasible.

The main contributions of this paper are: 

\begin{itemize}

\item \textbf{Hardware-aware deep networks for traffic classification:} 
To the best of our knowledge, this is the first work to design hardware-efficient deep networks for TC under strict, IoT-compatible constraints. HW-NAS specifically targets session-level classification and endows DNNs with hardware efficiency and satisfactory classification accuracy; previous approaches primarily focused on maximizing accuracy \cite{wang2017end,cui2019session,lu2021iclstm,only_header_cui2022only,seydali2023cbs,attention_spatio_hu2023network,zou2024novel_residual}.

\item \textbf{Header-level preprocessing and session length reduction:} This work extends existing studies by covering multiple header-level preprocessing methods, such as IP/\added{MAC addresses}, UDP padding, and anonymization, alongside session-length reductions, on the optimized model. Experimental evidence demonstrates that the model achieves stable performance, maintaining high accuracy with session lengths reduced by up to 75\%, while achieving greater efficiency for resource-constrained devices through these reductions. At the same time, excessive data obfuscation or aggressive reductions in session lengths can affect accuracy; hence, careful selection of configurations is required to avoid drops of up to 7\%. The paper gives practical guidelines for optimizing preprocessing and session length, providing a basis for future research on trading off accuracy and resource efficiency.

\item \textbf{Efficiency benchmarks:} Established benchmarks validate the HW-NAS-driven strategy against state-of-the-art (SOTA) methods. The proposed model yields substantial resource savings, with reductions of up to 444-fold in parameters, 312-fold in \added{FLOPs}, and 15-fold in maximum tensor size compared to baseline models. These reductions streamline session-level TC by reducing computational cost and memory usage.

\item \textbf{Comprehensive validation across tasks:} The HW-NAS-optimized model, tailored on the ISCX VPN-nonVPN dataset \cite{iscxvpnnonvpn}, demonstrates strong performance across diverse TC tasks. It achieves \deleted{96.59\%}\added{96.60\%} accuracy in the main VPN-nonVPN classification. Additionally, it delivers high accuracy in VPN differentiation \deleted{(99.95\%)}\added{(99.86\%)}, VPN-type \deleted{(99.19\%)}\added{(99.14\%)}, broader traffic categories \deleted{(96.94\%)}\added{(96.74\%)}, and session-level application identification \deleted{(94.61\%)}\added{(94.18\%)}, showing its flexibility and effectiveness in real-world scenarios.

\item \added{\textbf{Cross-dataset generalization:} The very same architecture, tailored on the ISCX dataset, proves able to achieve up to 99.98\% accuracy on the USTC-TFC2016~\cite{wang2017malware} and QUIC NetFlow~\cite{tong2018novel} datasets.}

\item \added{\textbf{On-device inference:} Efficiency gains from reduced session lengths also result in lower latency and energy consumption, enabling the model to run reliably on constrained IoT hardware; the quantized architecture operates in real time on Cortex-M4 and M7 microcontrollers, with latency under 115\,ms and energy below 29\,mJ.}

\item \added{\textbf{Code availability:} The proposed HW-NAS is available at \url{https://github.com/SEAlab-unige/ProtectIT_Unige}.}

\end{itemize}

\section{Related Work}

\subsection{Traditional Approaches}

Traditional TC methods such as port-based classification and DPI often fail on encrypted and obfuscated traffic.

Port-based classification associates traffic with specific services based on port numbers, which was suitable for older architectures but fails in modern networks due to dynamic ports and HTTP tunneling \cite{pacheco2018towards, survey_DONG2024128444}. Foundational studies indicated that default ports in Peer-to-Peer (P2P) protocols account for less than 70\% of actual traffic, thus confirming the increasing ineffectiveness of this method
\cite{ports_sen2004accurate,ports_moore2005toward}.

DPI identifies applications by detecting signatures within packet payloads \cite{pacheco2018towards, survey_DONG2024128444}. While effective on unencrypted traffic, it fails on encrypted flows where payloads are inaccessible~\cite{pacheco2018towards}, and it remains resource-intensive~\cite{survey_DONG2024128444}.

\subsection{Machine Learning Methods}

Classic machine learning (ML) approaches infer patterns from statistical features (e.g., packet size, flow duration, and arrival intervals) without examining packet payloads \cite{pacheco2018towards, survey_DONG2024128444}.

Various algorithms, including Na\"ive Bayes for real-time application classification, k-Nearest Neighbors (k-NN), and Decision Trees, have demonstrated effectiveness in identifying encrypted flows based on these features \cite{iscxvpnnonvpn}. Additionally, Random Forests (RF) have been successfully employed for TC in real-world specialized environments like software-defined networking (SDN) \cite{ml_rf_supervised_zhai2018random}.

Addressing the need for high-speed processing, \cite{akem2024encrypted} implemented a fully in-switch classification framework that executes RF inference at line rate within the data plane. Moreover, \cite{6476080} combined flow label propagation with compound classification to detect unknown flows and boost accuracy.

Some limitations affect traditional ML methods for TC. They rely on manual feature extraction, which is labor-intensive, and even strong models such as RF require frequent retraining to maintain satisfactory performance \cite{survey_DONG2024128444}.

DL models extract complex features from raw traffic, bypassing manual feature selection. While Convolutional Neural Networks (CNNs) and Recurrent Neural Networks (RNNs) achieve high accuracy, they are often computationally heavy, hindering deployment in resource-limited applications.
 
Several studies focused on session-level data. In \cite{wang2017end}, a 1D-CNN and 2D-CNN were evaluated on this task: the 1D-CNN scored an accuracy of 86.6\% and outperformed the 2D-CNN. The approach presented in \cite{wang2017malware} also used 1D-CNN for malware TC. \cite{he2020image} proposed an image-based approach that converted initial non-zero payload sessions into grayscale images for classification using CNNs. Although it achieved high F1 scores (97.73\% for conventional encrypted traffic and 99.55\% for VPN traffic), the coarse-grained handling of subflows disregarded relevant session information. The framework in \cite{wang2020encrypted} combined CNN and Stacked AutoEncoders (SAE) for encrypted traffic classification (ETC) by integrating trimmed raw traffic data and statistical features; the method reached 98\% F1-score, at the expense of additional time required for feature extraction.  \protect\added{Long Short-Term Memory (LSTM)} networks, often used to model sequential dependencies, were employed in the \protect\added{Inception-LSTM (ICLSTM)} model \cite{lu2021iclstm}, which converted traffic data into grayscale images and combined Inception modules with LSTM layers; it attained over 98\% accuracy but showed limitations in adapting to diverse real-world scenarios. The integration of LSTM, CNN, and a squeeze-and-excitation module to capture spatiotemporal features \cite{attention_spatio_hu2023network} achieved over 90\% accuracy on encrypted and unencrypted traffic via end-to-end representation learning. Focusing on attention mechanisms added computational overhead and limited scalability for resource-constrained environments.

At the packet level, \cite{lotfollahi2020deep} introduced \textit{Deep Packet}, which classified encrypted traffic using a combination of SAE and CNN. Packet-level approaches require large volumes of data, thus leading to high computational requirements and extended training times. The work in \cite{datanetwang2018} proposed \textit{DataNets}, which used \added{Multilayer Perceptron (MLP)}, SAE, and CNN models for the accurate categorization of encrypted traffic in smart home networks. Aggressive undersampling was used to address dataset imbalance, but it may affect the model's generalization performance. The authors in \cite{wang2020deep_packet_malicious}  proposed a deep hierarchical neural network for packet-level malicious traffic detection; a 1D-CNN for spatial feature extraction and \added{Gated Recurrent Unit (GRU)} for temporal features scored high accuracy levels across multiple intrusion detection datasets \cite{iscx2012,cicids2017}.

Several works integrated session and packet-level data.  Capsule Networks (CapsNet) \cite{cui2019session} yielded a session packet-based model for ETC. It prioritized classification performance at the expense of computational overhead. Likewise, the CBS model in \cite{seydali2023cbs} combined 1D-CNN, attention-based Bi-LSTM, and SAE for ETC at both session and packet levels by integrating spatial, temporal, and statistical features. CBS employed Generative Adversarial Networks (GANs) for data augmentation and class imbalance handling, resulting in improved performance. The computational requirements limited its use in offline processing and made the method unsuitable for real-time applications. \added{In~\cite{umair2021efficient}, the authors proposed a hybrid model that extracts flow-level statistical features and classifies them using a deep feedforward DNN followed by a maximum entropy classifier. Despite reporting 99.23\% accuracy, the approach relies on handcrafted features and static encoding, limiting its adaptability to raw or evolving traffic.}
Other works \cite{yao2019identification} applied attention-based LSTM and Hierarchical Attention Networks (HAN) to classify flows as time-series data. The attention mechanism captures key features, achieving 91.2\% accuracy, but with limited efficiency on constrained devices.

\subsection{Trends in Hardware-Aware NAS}
NAS methods improve model performance by tuning the network architecture on the target data.
Many approaches have been proposed in the literature \cite{1000papersNAS, chitty2023neural_ieee-access}, each improving architecture exploration in unique ways.

HW-NAS extends NAS by optimizing models for accuracy alongside hardware constraints, such as memory, latency, and computational efficiency \cite{hw-nas_survey_benmeziane2021comprehensive}\superadded{; this direction was established by works such as MnasNet \cite{tan2019mnasnet} and ProxylessNAS \cite{cai2018proxylessnas}, which integrated hardware objectives into the search process targeting mobile platforms}. \superadded{Frameworks like MCUNet \cite{lin2020mcunet} further specialized these techniques for microcontrollers by using system-algorithm co-design to fit deep models within strict SRAM and Flash constraints.} 
Additionally, studies such as \cite{ragusa2024combining, ragusa2024compression, garavagno2025searching} illustrated how HW-NAS could produce models tailored to efficient deployment on resource-limited platforms by matching network design with target hardware needs.

A recent work applied NAS to ETC \cite{nas_etc_malekghaini2023automl4etc}, although without strict IoT-centric constraints. \added{Other works have applied NAS to malware detection using proximal-iteration search~\cite{nas_zhang2023automatic}, and have shown that session-level ETC is feasible under hardware constraints~\cite{our_chehade2024tiny}. A HW-NAS strategy was also adopted in~\cite{chehade2025energy} for session-level inference on microcontrollers; however, the study did not account for how input data variation can affect efficiency and performance.}

NAS effectiveness relies on careful adaptation to the specific constraints and priorities of the target application domain. The research presented in this paper relies on HW-NAS to develop a hardware-efficient neural network for ETC. The target environment includes IoT applications, in which high accuracy and resource efficiency are of paramount importance.

\section{Preliminaries}

NAS automates neural network design by applying a search algorithm to identify optimal architectures within a predefined search space; this approach outperforms traditional handmade models \cite{chitty2023neural_ieee-access, nas_etc_malekghaini2023automl4etc} in terms of accuracy.

Conventional NAS algorithms aim to optimize computational performance but often overlook hardware constraints; as a result, the models they produce may lack the efficiency required for their deployment in resource-limited devices. HW-NAS overcomes this limitation by directly including hardware constraints in the search process; this makes it possible to envision deployment on a variety of application devices, such as microcontroller units (MCUs), field-programmable gate arrays (FPGAs), and cloud accelerators. The extended optimization problem takes into account multiple objectives and constraints, including accuracy, memory usage, computational complexity (measured in FLOPs), and energy efficiency \cite{1000papersNAS}.

The HW-NAS optimization problem is to identify an architecture \( a \in \mathcal{A} \) that minimizes the validation loss on a given dataset \( \mathcal{D} \). This objective can be formally defined as: 

\begin{mini}|s|
    {a \in \mathcal{A}} {\mathcal{L}_{\text{val}}(w^*(a), a)}
    {}{}
    \addConstraint{w^*(a) = \argmin_w \mathcal{L}_{\text{train}}(w, a)}
    \addConstraint{\psi(a, HW) < \text{Thr}}
    \label{eq:optnas}
\end{mini}
where \(\mathcal{A}\) denotes the search space containing all candidate architectures, \( \mathcal{L}_{\text{val}} \) represents the validation loss, and $\mathcal{L}_{\text{train}}$ is the training loss minimized to optimize the model weights \( w^*(a) \).  \( \psi(a, HW) \) is a function that measures the performance of the architecture with respect to hardware-specific metrics and \( \text{Thr} \) is a threshold value associated with these constraints. 

Three main phases characterize NAS, namely, defining the search space, selecting the search strategy, and evaluating the performance. The search space defines the set of admissible network structures (including layer configurations and optimization hyperparameters), which collectively determine the exploration boundaries for the NAS algorithm. The search strategy then controls how the NAS algorithm explores the architecture space. 
Finally, the performance evaluation tests the candidate architecture using a task-specific merit function, which adapts to the requirements of the application.

\section{Methodology: design of efficient network for traffic classification}

The method described in this paper provides a novel design pipeline to develop lightweight, optimized DNN architectures for session-level ETC, \added{with a focus on deployment on resource-constrained IoT devices.}
The workflow consists of three key stages: (1) \textbf{Data preprocessing} arranges raw traffic data for DL by converting sessions into a consistent input format; (2) in \textbf{HW-NAS execution}, the search algorithm explores the space of admissible architectures to find a high-performing model that also satisfies hardware constraints; finally, (3) the \textbf{Model selection and testing} first picks out the best-performing architecture, based on performance and constraints, then completes a rigorous testing process involving multiple tasks \added{and deployment settings, ensuring generalization and real-world feasibility.} % Figure \ref{fig:experiment_setup} outlines the overall design pipeline.

\subsection{Preprocessing}

The following steps implement the preprocessing phase.% (in the first block of Figure \ref{fig:experiment_setup}):

\textbf{Session extraction}: Raw traffic data is divided into sessions; each session consists of an ordered sequence of bidirectional packets exchanged between two endpoints, identified by IP addresses, port numbers, and protocol. Packet data is typically stored as raw bytes. %using Scapy, a Python library for network packet manipulation. 
    
\textbf{Data cleaning}: This step removes data-link layer information, including MAC addresses, and anonymizes IPs. This approach is frequently adopted in the literature to prevent overfitting; otherwise, some session-specific features could lead the model to memorize traffic patterns instead of learning generalization-relevant features \cite{lotfollahi2020deep,seydali2023cbs}.

\textbf{Filtering irrelevant data}: Packets without payloads (e.g., those featuring SYN, ACK, or FIN flags) and irrelevant DNS segments are discarded. Prior studies showed that those kinds of packets did not contribute meaningfully to TC and could actually add noise to the dataset. Filtering these packets ensures only meaningful traffic is used for training \cite{seydali2023cbs}.

\textbf{Session normalization}: In compliance with a method originally proposed in \cite{wang2017end} and \cite{wang2017malware}, each session is normalized to a uniform length of 784 bytes. This approach proved effective in handling varying session sizes while ensuring consistency for model input. Longer sessions are truncated, and shorter sessions are padded with null bytes to maintain a uniform length over all sessions.

\textbf{Data scaling}: This process prepares the data for subsequent input to the model, converting raw byte information into a standardized format suitable for neural network training. Each byte in the raw packet values forming session data is normalized into the range [0,1].

%\textbf{Storage}: Preprocessed session data is stored for subsequent training and evaluation steps in the HW-NAS process.

\subsection{Optimization Problem in HW-NAS} \label{optimization}

%To maximize generalization performance and meet hardware constraints, the HW-NAS design strategy implies an optimization problem, which can be formalized as follows:
The HW-NAS design strategy adopted in this paper casts architecture selection as a constrained optimization problem over candidate models $a \in \mathcal{A}$, defined as follows:

\begin{maxi}|s|
    {a \in \mathcal{A}} {\text{Accuracy}_{\text{val}}(w^*(a), a)}
    {}{}
    \addConstraint{w^*(a) = \argmin_w \mathcal{L}_{\text{train}}(w, a)}
    \addConstraint{|P(a)| < F_{\text{Th}}}
    \addConstraint{|T(a)| < R_{\text{Th}}}
    \addConstraint{\text{Flops}(a) < \text{Flops}_{\text{Th}}}
    \label{eq:optprob}
\end{maxi}

Hardware constraints are applied using thresholds $F_{\text{Th}}$, $R_{\text{Th}}$, and $\text{Flops}_{\text{Th}}$, which manage the critical resource limitations typically characterizing IoT devices. These thresholds stem from existing works, such as \cite{ragusa2024combining}.

The parameter-count threshold, $F_{\text{Th}}$, controls the total number of model parameters $P(a)$ of a candidate architecture and directly affects memory usage. Reducing that quantity minimizes the model’s memory footprint and helps fit the limited Flash memory available on edge devices. The threshold on the maximum tensor size, $R_{\text{Th}}$, bounds the peak intermediate activation size $T(a)$ during inference and ensures that it fits the available RAM. This is critical for session-level classification tasks, where efficient memory utilization is essential to avoid overflows during computation. 
Finally, the FLOPs threshold, $\text{Flops}_{\text{Th}}$, limits computational complexity by bounding the number of floating-point operations, which serves as a proxy for inference speed and energy consumption. This constraint is essential for maintaining real-time performance on devices that often lack dedicated \added{Graphics Processing Units (GPUs)}.

\subsection{Search Space Design}

The HW-NAS search space in this work is a block-wise structure relying on 1D CNNs, which efficiently capture dependencies in session-based traffic data. They extract localized structural patterns within packet content (e.g., payload and headers) and sequential trends across an entire session (e.g., bidirectional exchanges). Recurrent neural networks (RNNs), such as LSTMs and GRUs, can model sequential dependencies in session-based traffic data. At the same time, their step-by-step processing paradigm increases overhead and training time, making them unsuitable for real-time applications. Likewise, Transformers capture global relationships using self-attention but exhibit impractical memory and computational requirements for resource-constrained environments. In contrast, 1D CNNs can benefit from weight sharing and local connectivity to extract features efficiently, yielding a balance between computational cost and classification accuracy.

Each convolutional layer in the CNN block is defined by its kernel size, number of filters, padding, and stride. These parameters allow the model to flexibly adjust its receptive field and depth so that it can capture both detailed packet-level features and overall session-level patterns.
A series of batch normalization, ReLU activation, pooling (max or average, with adjustable pooling size), and dropout (with a configurable dropout rate) follow each convolutional layer. This improves feature representation while minimizing computational load. A Global Average Pooling (GAP) layer supports the final feature aggregation, which condenses spatial information into a compact representation and feeds a dense layer for the eventual classification outcome.

\subsection{Evolutionary Algorithm for NAS Search}

In the proposed HW-NAS framework, an evolutionary algorithm supports the exploration of architecture space.  Evolutionary algorithms proved to be highly effective toward that purpose \cite{1000papersNAS}. This approach is suitable to ensure that target architectures meet predefined hardware requirements \cite{ragusa2024combining}.

The search strategy starts from an initial (parent) architecture, $a_0$. A mutation function $R_m(a_p)$ then spawns candidate architectures (children). Mutations involve random modifications to the parent, such as adding new blocks with randomly generated hyperparameters, removing blocks to reduce complexity, or adjusting parameters within existing blocks (e.g., number of filters, kernel size). Each child is then evaluated against the predefined hardware constraints $\mathcal{H}$, and only those that satisfy them proceed to training.

This spawning/selection process iterates over a predefined number of generations $N_g$; each generation yields $N_c$ child architectures. If a child fails the hardware constraint check, further mutations are applied until the required number of admissible children is achieved. A maximum depth constraint limits the number of blocks within each architecture, ensuring that each child remains within a preset complexity level. 

\added{The search loop is outlined in Algorithm~\ref{alg:proc}.} After passing the hardware check, each child undergoes training on the training dataset $\mathcal{X_T}$, and is subsequently evaluated for validation accuracy on the validation set $\mathcal{X_V}$. The architecture with the highest validation accuracy becomes the new parent, $a_p$, for the next generation. \added{After all generations, the architecture with the highest validation score is selected.}

\begin{algorithm}[htbp]
\caption{\protect\added{Evolutionary NAS with hardware constraints.}}
\label{alg:proc}
\textbf{Inputs:}
\added{Training set $\mathcal{X_T} = \{X_i, y_i\}_{i=1...n}$, validation set $\mathcal{X_V} = \{X_i, y_i\}_{i=1...m}$, search space $\mathcal{A}$, initial parent $a_0$, mutation operator $R_m(\cdot)$, $N_c$ children per generation, number of generations $N_g$, evaluation function $E(a, \mathcal{X_V})$, hardware constraints $\mathcal{H}$.}

\vspace{1pt}

\textbf{Procedure:}
\begin{algorithmic}[1]
\State \added{Initialize parent architecture $a_p \leftarrow a_0$}
\State \added{Initialize best architecture $a^* \leftarrow a_p$}
\For {$g = 1$ to $N_g$}
    \For{$c = 1$ to $N_c$}
        \State \added{Generate child $a_c \leftarrow R_m(a_p)$}
        \If{\added{$a_c$ satisfies $\mathcal{H}$}}
            \State \added{Train $a_c$ on $\mathcal{X_T}$}
        \EndIf
    \EndFor
    \State \added{Select $a_p \leftarrow \arg\max\limits_{a_c} E(a_c, \mathcal{X_V})$}
    \State \added{Update $a^* \leftarrow \arg\max(E(a^*, \mathcal{X_V}), E(a_p, \mathcal{X_V}))$}
\EndFor
\State \added{\textbf{Return} $a^*$}
\end{algorithmic}
\end{algorithm}

\section{Experimental Setup}

\added{This section outlines the empirical setting for designing and evaluating efficient DNNs for session-level TC. It covers: (1) the encrypted traffic datasets; (2) the deployment-driven hardware constraints; and (3) the HW-NAS process for discovering models under resource limits.}

\subsection{Datasets and Tasks}
\subsubsection{Encrypted VPN Dataset}

The ISCX VPN-nonVPN dataset \cite{iscxvpnnonvpn}, consisting of approximately 30GB of traffic divided into 11 classes, is used in this study. It includes traffic for various applications in \protect\added{packet capture (PCAP)} format, labeled according to application and activity type. Table~\ref{tab:dataset} outlines the dataset structure, detailing traffic categories, encapsulation types (VPN and Non-VPN), and their respective proportions. Widely adopted in TC research \cite{wang2017end,lotfollahi2020deep,yao2019identification,lu2021iclstm,seydali2023cbs,zou2024novel_residual}, this dataset serves as the main benchmark for architecture optimization and supports the definition of various TC tasks.

\begin{table}[htbp]
\caption{ISCX VPN-nonVPN 2016 number of samples.}
\centering
\renewcommand{\arraystretch}{1.2}
\Large
\resizebox{\columnwidth}{!}{%
\begin{tabular}{
>{\centering\arraybackslash}c|
>{\centering\arraybackslash}c|
>{\centering\arraybackslash}c|
>{\centering\arraybackslash}c|
>{\centering\arraybackslash}c
}
\hline
\textbf{Category} &
\textbf{Application} &
\multicolumn{2}{c|}{\textbf{Encapsulation}} &
\makecell{\textbf{Category} \\ \textbf{Ratio (\%)}} \\
\cline{3-4}
 &  & \textbf{Non-VPN} & \textbf{VPN} &  \\
\hline
\multirow{2}{*}{Chat} &
aim chat, facebook, hangout, &
\multirow{2}{*}{7121} &
\multirow{2}{*}{4014} &
\multirow{2}{*}{25.98} \\
 & icq, skype & & & \\
\hline
Email & email, gmail & 5183 & 293 & 12.78 \\
\hline
Streaming & netflix, vimeo, youtube, spotify & 1464 & 474 & 4.52 \\
\hline
File Transfer & skype, scp, sftp, ftp & 1251 & 986 & 5.22 \\
\hline
\multirow{2}{*}{VoIP} &
skype, hangout, facebook, &
\multirow{2}{*}{5409} &
\multirow{2}{*}{16184} &
\multirow{2}{*}{50.39} \\
 & voipbuster & & & \\
\hline
P2P & bittorrent & - & 475 & 1.11 \\
\hline
\end{tabular}%
}
\label{tab:dataset}
\end{table}

These tasks span a range of complexities and objectives, as detailed in Table \ref{tab:experiment_descriptions}. The rows represent different tasks, while the columns specify the name of the experiment, its objective, and the number of output classes. VPN-NonVPN classifies traffic into 11 categories across both VPN and Non-VPN traffic. VPN-Diff performs binary classification to detect whether traffic is VPN or Non-VPN. VPN-Type and NonVPN-Type further classify traffic types within the VPN and Non-VPN groups into 6 and 5 categories, respectively. Traffic-Cat evaluates general traffic categorization across both VPN and Non-VPN traffic into 6 usage-based categories. Finally, App-ID identifies 15 specific network applications.

\begin{table}[htbp]
\centering
\caption{Overview of classification tasks.}
\label{tab:experiment_descriptions}
\renewcommand{\arraystretch}{1.2}
\small
\resizebox{\columnwidth}{!}{%
\begin{tabular}{
>{\centering\arraybackslash}c|
>{\centering\arraybackslash}c|
>{\centering\arraybackslash}c
}
\hline
\textbf{Experiment} &
\textbf{Description} &
\textbf{Classes} \\
\hline
VPN-NonVPN & VPN and Non-VPN traffic classification & 11 \\
\hline
VPN-Diff & Protocol encapsulation detection & 2 \\
\hline
VPN-Type & VPN traffic type classification & 6 \\
\hline
NonVPN-Type & Non-VPN traffic type classification & 5 \\
\hline
Traffic-Cat & Network usage categorization & 6 \\
\hline
App-ID & Application identification & 15 \\
\hline
\end{tabular}%
}
\end{table}

\subsubsection{\protect\added{Generalization Benchmarks}}

\protect\added{In addition to ISCX, we consider two public benchmarks, USTC-TFC2016~\cite{wang2017malware} and QUIC NetFlow~\cite{tong2018novel}, to evaluate generalization under different traffic distributions. Both datasets are provided in raw pcap format (approximately 4GB for USTC and 120GB for QUIC) and reflect distinct classification scenarios.}

\protect\added{%USTC-TFC2016 includes 20 traffic classes, divided into benign and malware categories. It spans a range of application protocols and communication behaviors, with both encrypted and unencrypted traffic.
USTC-TFC2016 includes 20 traffic classes across benign and malware categories, spanning diverse protocols and communication behaviors in both encrypted and unencrypted data.}

\protect\added{QUIC NetFlow focuses on encrypted services over the QUIC protocol, grouped into five usage categories: chat, voice, video, music, and file transfer.
It captures structural differences in encrypted traffic generated by the QUIC protocol, which encrypts transport-layer metadata over UDP.}

\protect\added{Table~\ref{tab:generalization_datasets} summarizes the key characteristics of both datasets. Each row specifies a traffic category, the corresponding application type, the number and share of samples, and the total number of classes per dataset.}

\begin{table}[htbp]
\centering
\caption{\protect\added{USTC-TFC2016 and QUIC NetFlow overview.}}
\label{tab:generalization_datasets}
\huge
\resizebox{\columnwidth}{!}{%
\begin{tabular}{
>{\centering\arraybackslash}c|
>{\centering\arraybackslash}c|
>{\centering\arraybackslash}c|
>{\centering\arraybackslash}c|
>{\centering\arraybackslash}c
}
\hline
\textbf{Dataset} &
\textbf{Category} &
\textbf{Applications} &
\makecell{\textbf{Total Samples} \\ \textbf{(\% of dataset)}} &
\textbf{\# Classes} \\
\hline

\multirow{3}{*}{\makecell{USTC-\\TFC2016}} 
  & Benign 
  & \makecell[l]{BitTorrent, Facetime, FTP, Gmail, MySQL,\\ Outlook, Skype, SMB, Weibo, WoW} 
  & \makecell{405,517 \\ \textbullet~81.0\%} 
  & \multirow{3}{*}{20} \\
\cline{2-4}
  & Malware 
  & \makecell[l]{Cridex, Geodo, Htbot, Miuref, Neris,\\ Nsis-ay, Shifu, Tinba, Virut, Zeus} 
  & \makecell{95,334 \\ \textbullet~19.0\%} 
  & \\
\hline
\hline

\multirow{7}{*}{\makecell{QUIC\\NetFlow}} 
  & Chat 
  & Google Hangouts Chat 
  & \makecell{14,528 \\ \textbullet~8.6\%} 
  & \multirow{7}{*}{5} \\
\cline{2-4}
  & Voice Call 
  & Google Hangouts VoIP 
  & \makecell{61,311 \\ \textbullet~36.3\%} 
  & \\
\cline{2-4}
  & File Transfer 
  & File Transfer Service 
  & \makecell{5,045 \\ \textbullet~3.0\%} 
  & \\
\cline{2-4}
  & Video Streaming 
  & YouTube 
  & \makecell{49,056 \\ \textbullet~29.1\%} 
  & \\
\cline{2-4}
  & Music Streaming 
  & Google Play Music 
  & \makecell{36,466 \\ \textbullet~21.6\%} 
  & \\
\hline
\end{tabular}%
}
\end{table}

\subsection{\protect\added{Target Hardware for Deployment}}

\added{We target two microcontroller platforms commonly used in embedded systems: the STM32F746G-DISCO and the Nucleo-F401RE. These boards represent distinct deployment scenarios for low-power TC at the edge: the STM32F7 suits moderately capable systems, while the Nucleo-F4 is a good example of a minimal IoT edge node.}

\protect\added{Table~\ref{tab:hardware_specs} reports key specifications for both targets. Rows list core type, Flash/RAM, clock speed, and power. These constraints set a reference for the deployment envelope and guide the design of lightweight models for on-device inference.}

\begin{table}[htbp]
\centering
\caption{\protect\added{Target microcontroller specifications.}}
\label{tab:hardware_specs}
\renewcommand{\arraystretch}{1.1}
\small
\resizebox{0.82\columnwidth}{!}{%
\begin{tabular}{
>{\centering\arraybackslash}c|
>{\centering\arraybackslash}c|
>{\centering\arraybackslash}c
}
\hline
\textbf{Specification} &
\textbf{STM32F746G-DISCO} &
\textbf{Nucleo-F401RE} \\
\hline
Core & Arm Cortex-M7 & Arm Cortex-M4 \\
\hline
Flash Memory & 1 MB & 512 KB \\
\hline
RAM & 340 KB & 96 KB \\
\hline
Clock Speed & 216 MHz & 84 MHz \\
\hline
Power Supply & 5V & 5V \\
\hline
\end{tabular}%
}
\end{table}

\subsection{\protect\added{HW-NAS Procedure}}

\added{The HW-NAS search is conducted on the VPN-NonVPN task from ISCX,} selected for its comprehensive nature and frequent use in prior research, which provides a solid foundation for comparisons with SOTA models. It includes both VPN and Non-VPN traffic, capturing shared structural patterns relevant to broader classification settings. The goal is to discover a network architecture that generalizes effectively across various tasks \added{and datasets}, without the need for task-specific designs.

%The search was run on a workstation with a Nvidia 2080 Ti GPU, receives preprocessed session data as input, which are extracted and prepared using Scapy, a Python library for packet manipulation. The HW-NAS codebase was built with Keras and TensorFlow.

We ran the search on a workstation with an NVIDIA 2080 Ti GPU, using session-level data preprocessed with Scapy, while the HW-NAS framework was built on Keras and TensorFlow.

The thresholds for memory usage and FLOPs \deleted{are defined based on}\added{have been initially set using} the minimum values observed among session-level TC models in prior works \added{as reference}. These metrics were calculated following details provided by each study and implemented in Keras to ensure consistency and accuracy. \protect\added{We found substantial headroom for improving efficiency. Since deployment targets both STM32F746G-DISCO and the more constrained Nucleo-F401RE, the thresholds were aligned with the latter’s resource limits (see Table~\ref{tab:hardware_specs}). Specifically, we set the maximum parameter count to 120K and the maximum tensor size to 22K, corresponding to 480\,KB Flash and 88\,KB RAM under 32-bit floating-point assumptions (float32). These estimates match the Nucleo’s 512\,KB Flash and 96\,KB RAM and enable compatibility checks during float32-based training and search. The FLOPs threshold was set to 11M based on the device’s 84\,MHz clock, which theoretically allows up to 84 million operations per second. This implies an upper bound of approximately 130\,ms for inference execution, excluding memory access overheads, which are not explicitly modeled in this estimate. This latency remains acceptable in session-level TC scenarios, where inference is invoked once per aggregated session, and it falls well within typical limits observed in edge-hosted analytics and network monitoring tasks, which tolerate response times between 100-250\,ms depending on application domain~\cite{jiang2018low_flops_latency}. Further compression (e.g., quantization or pruning) can provide additional margin at deployment. Empirical measurements on the target MCUs confirmed that latency remains within this limit, as reported in Section~\ref{deployment}.}

A validation set consisting of 20\% of the training data is created using a standard holdout method. Architectures are trained for up to 100 epochs, with an initial learning rate of $10^{-3}$, a batch size of 128, and a learning rate reduction triggered by a plateau in validation loss. Early stopping is applied using the validation loss as the stopping criterion. Networks are trained 3 times using a multi-start approach, \added{and the best validation accuracy is used to score each architecture.}

%The HW-NAS process spans 100 generations, with each generation containing a population of 10 candidate architectures (children). 
HW-NAS is conducted over 100 generations with 10 candidate architectures (children) per generation.
\added{On average, 16 children had to be generated per generation to obtain 10 admissible ones, as approximately 6 were discarded due to hardware constraint violations. Each child is derived through two random mutations applied to the parent, with mutation types drawn uniformly from block insertion, block removal, or block-level parameter modification (e.g., kernel size, filters). To prevent overcomplex architectures, the maximum depth is 5 blocks. This setup promotes structural diversity and reduces the risk of early convergence to suboptimal solutions.}

The search space includes filters ranging from 16 to 140, kernel sizes from 3 to 7, strides from 1 to 6, and dropout rates from 0.1 to 0.5. Pooling, if enabled, can be either max or average, with a pool size between 2 and 3. Padding is set as either “same” or “valid.” The best-performing model is selected based on validation accuracy. 

\begin{figure}[htbp]
    \centering
    \includegraphics[width=\columnwidth]{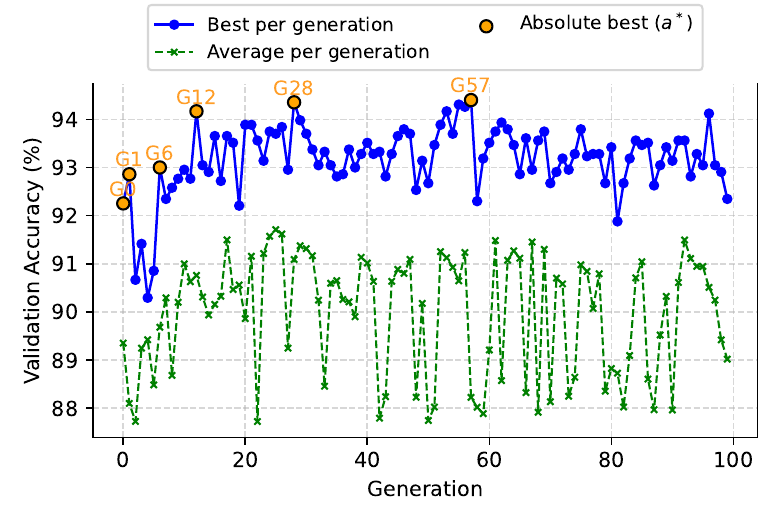}
    \caption{\protect\added{Validation accuracy over HW-NAS generations.}}

    \label{fig:nas_convergence}
\end{figure}

\added{Figure~\ref{fig:nas_convergence} shows the validation accuracy of the best candidate per generation (solid line), along with the average accuracy of all admissible children (dashed line). Orange markers indicate generations where a new absolute best model was found (e.g., G0, G1, G6, G12, G28, G57); no further improvements occurred beyond G57, indicating that the search converged early under the imposed hardware constraints. The average curve complements this view by capturing the overall performance trend of each generation beyond the individual best models.}

\deleted{The model generated by the NAS}\added{The best model found by HW-NAS} is retrained \deleted{across all experiments}\added{on all tasks} in Table \ref{tab:experiment_descriptions} \protect\added{and evaluated on the external datasets in Table~\ref{tab:generalization_datasets}} to assess its generalization performance. %Each experiment is trained for up to 200 epochs with early stopping.
\added{For these retraining experiments, we performed 10 independent runs and reported the sample standard deviation of accuracy to capture performance variability.} 
Each experiment is trained with early stopping, \added{using a maximum of 50 to 200 epochs depending on the task and dataset, with higher limits applied when convergence required more iterations.}

\section{Results and Evaluation} \label{results}

\subsection{HW-NAS-Optimized Architecture}

The best architecture identified by HW-NAS is shown in Table \ref{tab:custom_cnn_model}. %The table rows correspond to the sequential layers of the CNN, while the columns specify each layer's attributes:
Rows list the sequential CNN layers, and columns report the corresponding layer attributes: layer type (e.g., \added{one-dimensional convolution (Conv1D)}, \added{average pooling (AvgPool1D)}, \added{max pooling (MaxPool1D)}, or Dense), input dimensions (Input Dim), number of filters or units (Filt/Units), kernel or pooling size (Ker/Pool), stride (Str), and padding (Pad).

\begin{table}[htbp]
\centering
\caption{Best CNN architecture identified by HW-NAS.}
\label{tab:custom_cnn_model}
\renewcommand{\arraystretch}{1.1}
\large
\resizebox{\columnwidth}{!}{%
\begin{tabular}{
>{\centering\arraybackslash}c|
>{\centering\arraybackslash}c|
>{\centering\arraybackslash}c|
>{\centering\arraybackslash}c|
>{\centering\arraybackslash}c|
>{\centering\arraybackslash}c|
>{\centering\arraybackslash}c
}
\hline
\textbf{\#} &
\textbf{Layer} &
\textbf{Input Dim} &
\textbf{Filt/Units} &
\textbf{Ker/Pool} &
\textbf{Str} &
\textbf{Pad} \\
\hline
1 & Conv1D + ReLU & 784 x 1 & 129 & 7 & 5 & valid \\
\hline
2 & Conv1D + ReLU & 156 x 129 & 110 & 4 & 2 & valid \\
\hline
3 & AvgPool1D & 77 x 110 & - & 3 & 2 & same \\
\hline
4 & Conv1D + ReLU & 39 x 110 & 38 & 7 & 2 & valid \\
\hline
5 & MaxPool1D & 17 x 38 & - & 2 & 2 & same \\
\hline
6 & GAP & 9 x 38 & - & - & - & - \\
\hline
7 & Dense + Softmax & 38 & 11 & - & - & - \\
\hline
\end{tabular}%
}
\end{table}

A key aspect is the progressive reduction of filters across convolutional layers, which allows the model to balance feature extraction and computational cost. Kernel size and stride vary adaptively: early layers use larger receptive fields to capture broad patterns, while deeper layers focus on finer details.
This design achieves both accuracy and hardware efficiency, with only 88.26K parameters, a maximum tensor size of 20.12K, and 10.08M FLOPs. 
\added{These values are comfortably below the thresholds set by the Nucleo-F401RE and STM32F7 platforms, confirming on-device deployability; they also indicate likely compatibility with other microcontrollers and low-power edge devices available on the market with similar or higher resource capacity.}

\added{In the following comparisons, we assess if SOTA models meet RAM and Flash limits using float32 parameter counts and max tensor sizes as proxies. Models above these limits may run on more capable platforms, but require structural reduction to operate on the tested low-power MCU class \superadded{(e.g., redesign or compression)}. Furthermore, when a model fits within the memory constraints but exceeds the proposed model’s 10.08M FLOPs, one should expect both latency and energy consumption to increase on comparable devices.
}

\subsection{\protect\added{VPN-NonVPN Task: Performance and Efficiency}}

%This subsection presents the results of the VPN-NonVPN experiment at the session level, i.e., the scenario of interest for this research.
This subsection reports the session-level results of the VPN-NonVPN experiment, the target scenario of this study.
% highlighting the efficiency and competitive performance of the model.

\begin{figure}[htbp]
    \centering
    \includegraphics[width=\columnwidth]{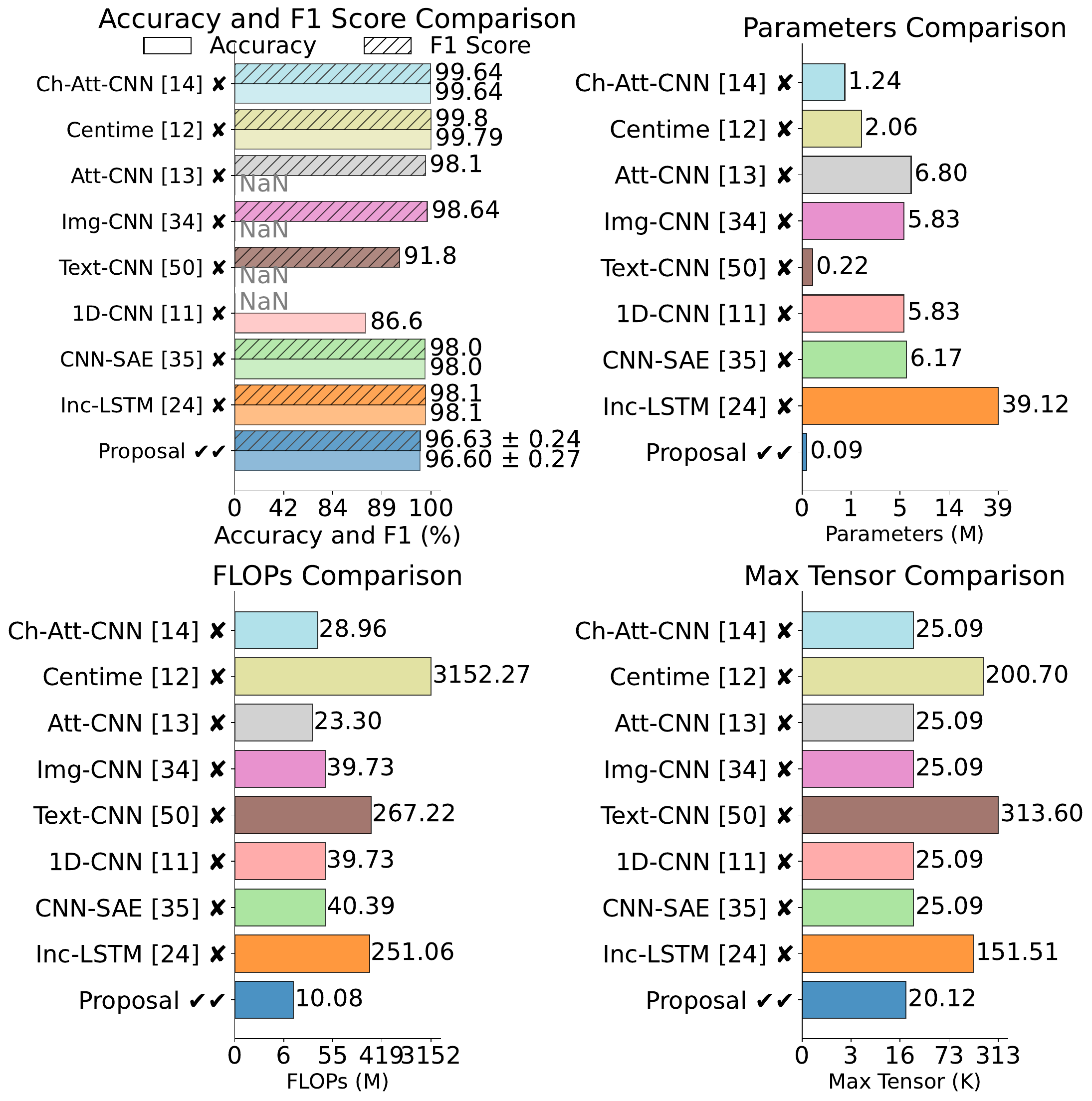}
        \caption{Comparison of session-level methods based on accuracy, F1 score, parameters, max tensor, and FLOPs.
        }
    \label{fig:comparison_session_accuracy_params_flops_maxtensor}
\end{figure}

Figure \ref{fig:comparison_session_accuracy_params_flops_maxtensor} compares accuracy, F1 score, parameters, FLOPs, and maximum tensor size across SOTA models. \added{The proposed model is annotated with its standard deviation across ten independent training runs. A single check mark~(\checkmark{}) indicates compatibility with STM32F7-class MCUs; a double check mark~(\checkmark\checkmark{}) denotes compatibility with both STM32F7 and the more constrained F401RE platform; a cross~($\times$) marks cases where deployment is not feasible. No other model in this session-level comparison meets both constraints, underscoring the distinctive efficiency of the proposed architecture.}

The proposed model achieves an accuracy of \deleted{96.59\%}\added{96.60\%} and an F1 score of \deleted{96.54\%}\added{96.63\%}, which are slightly lower than some of the most resource-intensive models in the SOTA. For example, \cite{centime_maonan2021centime} and \cite{zou2024novel_residual} achieve accuracy and F1 scores exceeding 99\% but require 2.06M and 1.24M parameters, respectively, compared to the proposed model's 0.09M with a reduction of up to 22 times. Their FLOPs requirements (3152.27M and 28.96M) surpass the proposed model's (10.08M) by up to 312 times and nearly 2.9 times, respectively. The maximum tensor size in certain works, such as 313.60K in \cite{song2019encrypted}, further underscores the efficiency of the proposed model, which operates with a tensor size of only 20.12K, achieving a reduction of over 15 times. \cite{lu2021iclstm} achieves an accuracy of 98.10\%, requiring an increment of 435, 24.9, and 7.5 times of the three hardware estimators. Similarly, \cite{wang2020encrypted} achieves 98.00\% accuracy at the expense of 68, 4, and 1.25 times increment. 
\deleted{These results clarify the capability of the proposed approach to balance computing requirements and generalization performance when compared with existing solutions.}\added{These results highlight the proposed model’s ability to deliver competitive performance compared to SOTA, while remaining within the strict memory and compute constraints of microcontroller-class devices.}

\begin{figure}[htbp]
    \centering
    \includegraphics[width=\columnwidth]{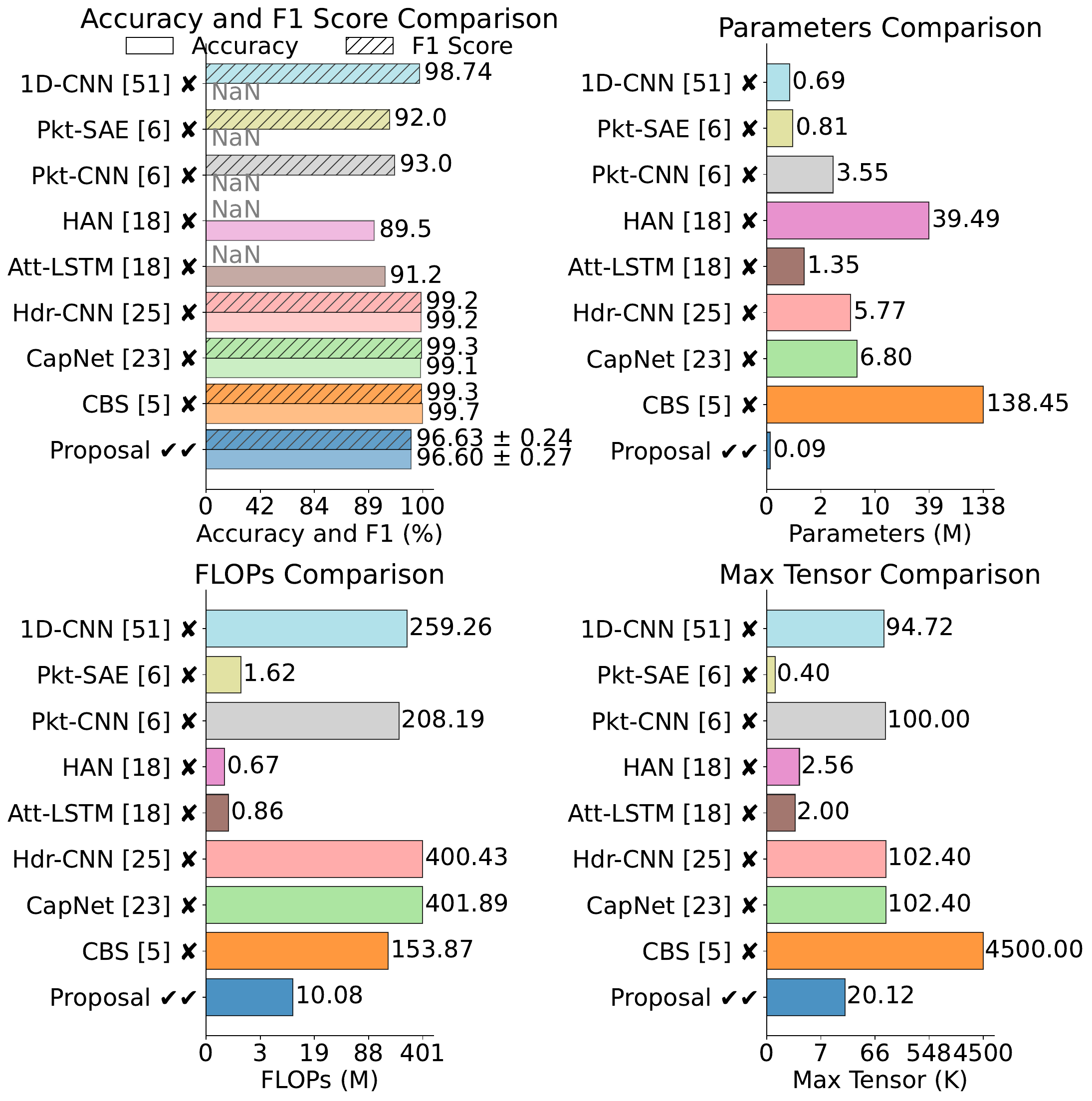}
    \caption{Comparison of non-session methods based on accuracy, parameters, max tensor, and FLOPs.
    }
    
    \label{fig:comparison_non_session_accuracy_params_flops_maxtensor}
\end{figure}

Figure~\ref{fig:comparison_non_session_accuracy_params_flops_maxtensor} extends the analysis to non-session methods, which include packet-level~\cite{lotfollahi2020deep,malicious_packets_zhou2023identification}, flow-based~\cite{yao2019identification,only_header_cui2022only}, and hybrid~\cite{seydali2023cbs,cui2019session} input approaches. These differ in format, preprocessing, and problem framing, making direct comparisons inherently complex. The figure mirrors the structure of Figure~\ref{fig:comparison_session_accuracy_params_flops_maxtensor}, reporting accuracy, F1 score, parameters, FLOPs, and maximum tensor size. \added{The proposed model is annotated with its standard deviation over ten independent runs. It is also the only configuration that fits within both STM32F7 and F401RE constraints, as denoted by the \checkmark\checkmark{} symbol; a cross~($\times$) marks configurations where deployment is not feasible.}

Packet-level methods, such as the SAE of \cite{lotfollahi2020deep}, proved efficient, requiring 0.81M parameters and 1.62M FLOPs in their most efficient setup. \deleted{This comes at the cost of reduced F1 score (92.00\%), preprocessing overhead, and undersampling, which limits the generalizability to complex scenarios.}\added{However, its parameter count still exceeds the available flash memory of our target MCUs unless structural reduction is applied. This setup achieves an F1 score of 92.00\%.} The CNN-based approach of \cite{lotfollahi2020deep} is a larger model, demanding 3.55M parameters and 208.19M FLOPs (F1 score 93.00\%). Similarly, \cite{malicious_packets_zhou2023identification} incurs high hardware costs, requiring 0.69M parameters and 259.26M FLOPs.

Flow-based methods \cite{yao2019identification}, which process flows as time series with predefined packet counts and lengths, show similar trade-offs. \deleted{Despite their hardware efficiency, with 1.35M parameters and 0.86M FLOPs in one configuration and 39.49M parameters and 0.67M FLOPs in another, their accuracy (91.20\% and 89.50\%, respectively) remains significantly below the proposed model's 96.59\%.}\added{While their configurations are relatively efficient in terms of FLOPs, their parameter counts (1.35M and 39.49M, respectively) surpass the limits of the selected target devices. Their reported accuracy (91.20\% and 89.50\%, respectively) remains lower than that of the proposed model (96.60\%).} Similarly, \cite{only_header_cui2022only}, which focuses on flow headers for classification, achieves an accuracy of 99.20\% but requires 5.77M parameters and 400.43M FLOPs, exceeding the target resource envelope.

Hybrid input methods, like \cite{seydali2023cbs}, which combines raw packet data with hand-crafted session features, and \cite{cui2019session}, which segments sessions based on thresholds, achieve superior accuracy (99.70\% and 99.10\%, respectively). These methods have high requirements, with \cite{seydali2023cbs} demanding 138.45M parameters, 153.87M FLOPs, and a max tensor size of 4500.00K, and \cite{cui2019session} requiring 6.80M parameters and 401.89M FLOPs.

This analysis highlights the balance of competitive accuracy and remarkable efficiency achieved by the proposed model, solidifying its position as a reference point for hardware-aware TC, even when compared across different problem formulations and input setups. \added{While some other methods report higher accuracy, they do so at the cost of exceeding the memory or compute constraints set by the target devices selected in this work.}

%\subsection{\protect\deleted{Generalization Across Traffic Classification Tasks}}
\subsection{\protect\added{Evaluation on Multiple Traffic Classification Tasks}}
Table \ref{tab:comparison_experiment_metrics} includes accuracy and hardware efficiency metrics for different methods in diverse TC tasks. Each row represents a method (Meth.) and its corresponding input type, while columns detail accuracy for VPN-Diff (VPN-D.), VPN-Type (VPN-T.), NonVPN-Type (NVPN-T.), and Traffic-Cat (T-Cat.) tasks, along with parameters (Par.), max tensor size (M-Tens.), and FLOPs as hardware measures. \added{Proposed model results are reported as mean ± standard deviation over ten runs.}

\begin{table}[htbp]
\centering
\caption{Accuracy and hardware for multiple experiments.}
\label{tab:comparison_experiment_metrics}
\renewcommand{\arraystretch}{1.1}
\Huge
\resizebox{\columnwidth}{!}{%
\begin{tabular}{
>{\centering\arraybackslash}c|
>{\centering\arraybackslash}c|
>{\centering\arraybackslash}c|
>{\centering\arraybackslash}c|
>{\centering\arraybackslash}c|
>{\centering\arraybackslash}c|
>{\centering\arraybackslash}c|
>{\centering\arraybackslash}c|
>{\centering\arraybackslash}c
}
\hline
\textbf{Meth.} &
\textbf{Input} &
\makecell{\textbf{VPN-D.} \\ (\%)} &
\makecell{\textbf{VPN-T.} \\ (\%)} &
\makecell{\textbf{NVPN-T.} \\ (\%)} &
\makecell{\textbf{T-Cat.} \\ (\%)} &
\makecell{\textbf{Par.} \\ (M)} &
\makecell{\textbf{M-Tens.} \\ (K)} &
\makecell{\textbf{FLOPs} \\ (M)} \\
\hline
\textbf{Prop.} & \textbf{Sess.} 
& \makecell{\added{\textbf{99.86}} \\ \Huge$\pm$\added{\textbf{0.08}}} 
& \makecell{\added{\textbf{99.14}} \\ \Huge$\pm$\added{\textbf{0.17}}} 
& \makecell{\added{\textbf{94.04}} \\ \Huge$\pm$\added{\textbf{0.70}}} 
& \makecell{\added{\textbf{96.74}} \\ \Huge$\pm$\added{\textbf{0.57}}} 
& \textbf{0.09} & \textbf{20.12} & \textbf{10.08} \\
\hline
\cite{lu2021iclstm} & Sess. & 100 & 99.00 & 98.70 & 98.20 & 39.12 & 151.51 & 251.06 \\
%\hline
\cite{wang2017end} & Sess. & 99.90 & 98.30 & 81.70 & - & 5.83 & 25.09 & 39.73 \\
%\hline
\cite{seydali2023cbs} & Hybrid & 99.82 & 99.38 & 99.45 & - & 138.45 & 4500.00 & 153.87 \\
%\hline
\cite{yao2019identification} & Flows & 99.97 & 94.80 & 89.30 & - & 1.35 & 2.00 & 0.86 \\
%\hline
\cite{yao2019identification} & Flows & 99.50 & 92.90 & 85.10 & - & 39.49 & 2.56 & 0.67 \\
\hline
\end{tabular}%
}
\end{table}

The proposed model (Prop.) achieves near-perfect accuracy \deleted{(99.95\%)}\added{(99.86\%)} in VPN-Diff, matching \cite{lu2021iclstm} and surpassing \cite{seydali2023cbs} (99.82\%). In VPN-Type classification, it achieves \deleted{99.19\%}\added{99.14\%}, outperforming \cite{yao2019identification} (94.80\% and 92.90\%) and \cite{wang2017end} (98.30\%).

In the NonVPN-Type task, the model delivers strong performance \deleted{(94.17\%)}\added{(94.04\%)}, exceeding \cite{yao2019identification} (89.30\% and 85.10\%) and \cite{wang2017end} (81.70\%). For Traffic-Cat, a challenging task addressed only by \cite{lu2021iclstm}, the proposal scores \deleted{96.94\%}\added{96.74\%}. Although \cite{lu2021iclstm} obtains slightly higher accuracy (98.20\%), the proposal maintains competitive performance while being resource-efficient.

The results validate the proposal across a broad spectrum of tasks, achieving high accuracy and often outperforming SOTA models. As the architecture was optimized for the VPN-NonVPN classification task using HW-NAS, the proposed results can be considered a worst-case analysis.

\subsection{Performance on Application Identification (App-ID)}

The App-ID experiment assesses the ability of the proposed session-level model to classify specific applications, a task traditionally dominated by packet-level approaches and hybrid methods~\cite{seydali2023cbs}. Table \ref{tab:experiment6_metrics} presents accuracy (Acc.), F1 score (F1), and hardware efficiency metrics: parameters (Par.), maximum tensor size (M-Tens.), and FLOPs (M) for the App-ID experiment. Each row represents a method (Meth.) and its corresponding input type. \added{Scores for our model include standard deviation across ten independent runs.}

\begin{table}[htbp]
\centering
\caption{Performance for the App-ID experiment.}
\label{tab:experiment6_metrics}
\renewcommand{\arraystretch}{1.05}
\scriptsize
\resizebox{\columnwidth}{!}{%
\begin{tabular}{
>{\centering\arraybackslash}c|
>{\centering\arraybackslash}c|
>{\centering\arraybackslash}c|
>{\centering\arraybackslash}c|
>{\centering\arraybackslash}c|
>{\centering\arraybackslash}c|
>{\centering\arraybackslash}c
}
\hline
\textbf{Meth.} &
\textbf{Input} &
\makecell{\textbf{Acc.} \\ (\%)} &
\makecell{\textbf{F1} \\ (\%)} &
\makecell{\textbf{Par.} \\ (M)} &
\makecell{\textbf{M-Tens.} \\ (K)} &
\makecell{\textbf{FLOPs} \\ (M)} \\
\hline
\textbf{Prop.} &
\textbf{Sess.} &
\makecell{\added{\textbf{94.18}} \\ \scriptsize$\pm$\added{\textbf{0.54}}} &
\makecell{\added{\textbf{94.42}} \\ \scriptsize$\pm$\added{\textbf{0.51}}} &
\textbf{0.09} &
\textbf{20.12} &
\textbf{10.08} \\
\hline
\cite{seydali2023cbs} & Hybrid & 99.67 & 99.51 & 138.45 & 4500.00 & 153.87 \\
%\hline
\cite{lotfollahi2020deep} & Packets & 98.00 & 98.00 & 3.55 & 100.00 & 208.19 \\
%\hline
\cite{lotfollahi2020deep} & Packets & - & 95.00 & 0.81 & 0.40 & 1.62 \\
\hline
\end{tabular}%
}
\end{table}

The proposal (Prop.) achieves \deleted{94.61\%}\added{94.18\%} accuracy and \deleted{94.59\%}\added{94.42\%} F1 score with reduced hardware cost. \cite{lotfollahi2020deep} presents two packet-level models: one with higher accuracy (98.00\%) but greater resource use, and another reaching 95.00\% F1 with 0.81M parameters and 1.62M FLOPs. \added{The latter model would require a Flash memory larger than the one available in the target devices.} \deleted{Packet-level methods, however, must process a larger volume of data due to the granularity of individual packets, often requiring undersampling to manage data size.}\added{Packet-level models operate at finer granularity, which leads to larger training volumes and more frequent inference. While this may enable fast per-packet predictions and strong App-ID accuracy, it also results in higher cumulative processing overhead. These structural differences inherently affect both performance and resource use, and should be considered when interpreting the comparison with session-level methods.}

CBS~\cite{seydali2023cbs} achieves the highest accuracy (99.67\%) and F1 score (99.51\%) but requires higher resource consumption. \added{Moreover, it was explicitly designed for offline use and is documented as unsuitable for real-time scenarios, which limits its applicability to embedded deployment.}

These results confirm the feasibility of session-level models for application classification, achieving competitive performance with substantially lower hardware cost. \added{Despite the inherent granularity differences, the session-level approach remains a practical and scalable option for constrained platforms.}

\subsection{\protect\added{Cross-Dataset Evaluation}}

\added{This subsection evaluates how well the architecture discovered by HW-NAS on ISCX generalizes to other traffic distributions. We report results on USTC-TFC2016 and QUIC NetFlow. The proposed model is re-evaluated on these datasets using the same architecture found during search (see Table~\ref{tab:custom_cnn_model}).}

\begin{figure}[htbp]
    \centering
    \includegraphics[width=\columnwidth]{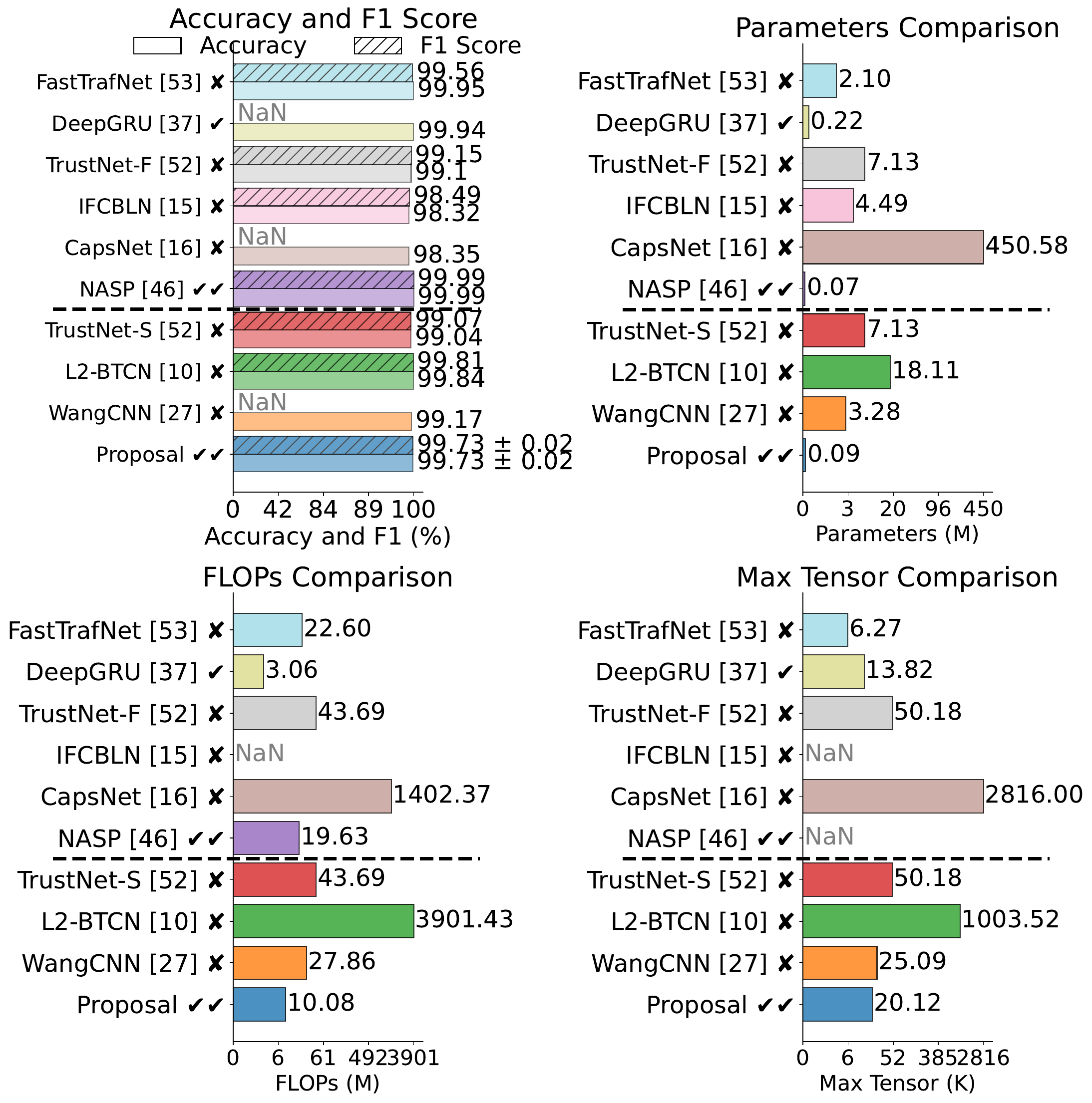}
    \caption{\protect\added{Results on the USTC-TFC2016 dataset.}}
    \label{fig:ustc_comparison}
\end{figure}

\added{Figure~\ref{fig:ustc_comparison} shows results on the USTC-TFC2016 dataset. The comparison includes session-based models~\cite{wang2017malware,li2024l2,li2025trustworthy}, flow-based models~\cite{nas_zhang2023automatic,xu2024cascaded,yao2019capsule,li2025trustworthy}, and packet-based models~\cite{wang2020deep_packet_malicious,yu2025model}. A horizontal line in the figure separates the session-based approaches (shown below), which are the focus of this study, from the rest. Reported scores for the proposed model include standard deviation over ten independent runs. Device compatibility follows the same notation as previous figures: \checkmark{} for STM32F7, \checkmark\checkmark{} for both STM32F7 and F401RE, and $\times$ where deployment is not feasible.}

\added{The proposed model achieves 99.73\% accuracy and 99.73\% F1 score with 0.09M parameters, 10.08M FLOPs, and a 20.12K max tensor. Although optimized on ISCX, it maintains strong performance on USTC with minimal hardware cost.}

\added{Among session-level models,~\cite{li2024l2} reports the highest accuracy (99.84\%) but at the cost of 18.11M parameters and over 3900M FLOPs. The session-based variant of~\cite{li2025trustworthy} also incurs high compute demand (7.13M parameters, 43.69M FLOPs).~\cite{wang2017malware} achieves similar accuracy (99.17\%) with fewer resources than the others in its category, though its compute and memory footprint remains higher than ours. None of these baselines are deployable on the tested MCUs due to excessive parameter and tensor sizes, making our model the only viable session-level solution in constrained settings.}

\added{The NAS-based flow model from~\cite{nas_zhang2023automatic}, tailored for USTC, reaches 99.99\% accuracy with slightly fewer parameters (0.07M) but nearly doubles the FLOPs (19.63M). Its max tensor is unreported, but we conservatively assume it fits within memory limits. However, its higher compute load suggests greater energy use and latency than ours on similar hardware. Other flow-level models~\cite{yao2019capsule,xu2024cascaded,li2025trustworthy} are far more complex, with parameter counts from 4.49M to 450.58M.}

\added{The model presented in \cite{wang2020deep_packet_malicious} achieves 99.94\% accuracy with 0.22M parameters, 3.06M FLOPs, and a 13.82K tensor, making it compatible with STM32F7-class devices. However, it exceeds the Flash constraints of F401RE and operates at packet-level granularity, requiring inference at finer time scales. This incurs higher cumulative overhead than our session-level model, which classifies entire sessions in a single pass. \cite{yu2025model} combines high accuracy (99.95\%) with the smallest tensor (6.27K), yet its 2.10M parameter count far exceeds the Flash capacity of our target devices.}

\added{Table~\ref{tab:quic_netflow_metrics} presents classification results on the QUIC NetFlow dataset. Each row corresponds to a method, while columns report input type, accuracy, and F1 score. Hardware metrics for existing methods are omitted due to missing implementation details. Compared approaches rely on flow-level features~\cite{akem2024encrypted}, handcrafted signals~\cite{dillbary2024hidden}, or hybrid pipelines combining raw bytes with flow features~\cite{tong2018novel}.}

\begin{table}[htbp]
\centering
\caption{\protect\added{QUIC NetFlow: classification performance.}}
\label{tab:quic_netflow_metrics}
\renewcommand{\arraystretch}{1.05}
\scriptsize
\resizebox{0.75\columnwidth}{!}{%
\begin{tabular}{
>{\centering\arraybackslash}c|
>{\centering\arraybackslash}c|
>{\centering\arraybackslash}c|
>{\centering\arraybackslash}c
}
\hline
\textbf{Method} &
\textbf{Input} &
\makecell{\textbf{Acc.} \\ (\%)} &
\makecell{\textbf{F1} \\ (\%)} \\
\hline
\textbf{Proposal} &
\textbf{Sess.} &
\makecell{\textbf{99.98} \\ \scriptsize$\pm$\added{\textbf{0.01}}} &
\makecell{\textbf{99.98} \\ \scriptsize$\pm$\added{\textbf{0.01}}} \\
\hline
CNN + RF\cite{tong2018novel} & Hybrid & 99.00 & 99.00 \\
%\hline
DWT-TC\cite{dillbary2024hidden} & Features & 93.70 & 93.60 \\
%\hline
STFT-TC\cite{dillbary2024hidden} & Features & 95.50 & 95.40 \\
%\hline
RF\cite{akem2024encrypted} & Features & 95.30 & 95.30 \\
\hline
\end{tabular}%
}
\end{table}

\added{Our model attains the highest accuracy (99.98\%) and F1 score (99.98\%) with minimal complexity (0.09M parameters, 10.08M FLOPs); results are averaged over ten runs. CNN+RF~\cite{tong2018novel} performs competitively (99.00\%) but uses a two-stage pipeline with limited deployment suitability. Feature-based methods~\cite{dillbary2024hidden,akem2024encrypted} lag by 4 to 6 points. These results show that a compact session-based model using raw bytes can outperform protocol-specific and feature-heavy pipelines, even in specialized QUIC traffic scenarios.}

\added{Overall, these findings confirm the model’s ability to generalize across encrypted traffic datasets with differing characteristics, while maintaining its efficiency advantage.}

\section{Evaluation and Analysis}
\subsection{Preprocessing Analysis}

Header-level preprocessing is critical for balancing privacy and the retention of key structural features essential for TC. Packet headers carry structural and protocol information that is particularly relevant for ETC, where the payload is obfuscated and contributes little to feature extraction. This study modifies specific header fields, commonly targeted in prior works, to evaluate their impact on classification performance in the VPN-NonVPN task\deleted{, which is a representative test case} \added{from ISCX, which serves as the reference benchmark for analysis}. The dataset predominantly includes TCP and UDP protocols, where headers follow a layered hierarchy: the Ethernet layer (14 bytes), containing source and destination MAC addresses (6 bytes each) and EtherType (2 bytes); the IP layer, which includes source and destination IP addresses (4 bytes each); and the transport layer, which adds source and destination ports (2 bytes each) and protocol headers, with UDP headers being 8 bytes and TCP headers 20 bytes. These modifiable fields collectively account for approximately 26 to 38 bytes per packet, out of the standardized session input size of 784 bytes.

Table \ref{tab:preprocessing_strategies} evaluates the effects of 24 preprocessing strategies (Strat.), i.e., all possible combinations of preprocessing choices for header fields, on the model derived from the NAS. Each row represents a strategy, with columns describing applied steps such as removing the Ethernet layer (Eth. Rem.), anonymizing (Anon.; replacing field values with hashed pseudonyms), zeroing (Zero; setting values to 0) MAC and IP addresses, zeroing protocol-related fields like ports (Port Zero), and applying UDP padding (UDP Pad.) to align UDP and TCP segments. A checkmark (\checkmark) indicates that a step was applied. The final column shows the accuracy (\%; Acc.). It should be noted that zeroing enhances privacy by replacing original values with zeros, which removes meaningful variability and makes it harder for the model to infer patterns.

\begin{table}[htbp]
\centering
\caption{Effect of preprocessing on model performance.}
\label{tab:preprocessing_strategies}
\renewcommand{\arraystretch}{1.1}
\resizebox{\columnwidth}{!}{%
\begin{tabular}{
>{\centering\arraybackslash}c|
>{\centering\arraybackslash}c|
>{\centering\arraybackslash}c|
>{\centering\arraybackslash}c|
>{\centering\arraybackslash}c|
>{\centering\arraybackslash}c|
>{\centering\arraybackslash}c|
>{\centering\arraybackslash}c|
>{\centering\arraybackslash}c
}
\hline
\textbf{Strat.} &
\makecell{\textbf{Eth.} \\ \textbf{Rem.}} &
\makecell{\textbf{MAC} \\ \textbf{Anon.}} &
\makecell{\textbf{MAC} \\ \textbf{Zero}} &
\makecell{\textbf{IP} \\ \textbf{Anon.}} &
\makecell{\textbf{IP} \\ \textbf{Zero}} &
\makecell{\textbf{Port} \\ \textbf{Zero}} &
\makecell{\textbf{UDP} \\ \textbf{Pad.}} &
\makecell{\textbf{Acc.} \\ \textbf{(\%)}} \\
\hline
1  & \checkmark &  &  & \checkmark &  &  & \checkmark & 95.35 \\ \hline
2  & \checkmark &  &  & \checkmark &  &  &  & 96.59 \\ \hline
3  & \checkmark &  &  & \checkmark &  & \checkmark & \checkmark & 95.80 \\ \hline
4  & \checkmark &  &  & \checkmark &  & \checkmark &  & 96.12 \\ \hline
5  & \checkmark &  &  &  & \checkmark &  & \checkmark & 95.56 \\ \hline
6  & \checkmark &  &  &  & \checkmark &  &  & 95.38 \\ \hline
7  & \checkmark &  &  &  & \checkmark & \checkmark & \checkmark & 95.12 \\ \hline
8  & \checkmark &  &  &  & \checkmark & \checkmark &  & 95.00 \\ \hline
9  &  & \checkmark &  & \checkmark &  &  & \checkmark & 94.79 \\ \hline
10 &  & \checkmark &  & \checkmark &  &  &  & 95.42 \\ \hline
11 &  & \checkmark &  & \checkmark &  & \checkmark & \checkmark & 95.61 \\ \hline
12 &  & \checkmark &  & \checkmark &  & \checkmark &  & 95.82 \\ \hline
13 &  & \checkmark &  &  & \checkmark &  & \checkmark & 95.45 \\ \hline
14 &  & \checkmark &  &  & \checkmark &  &  & 94.70 \\ \hline
15 &  & \checkmark &  &  & \checkmark & \checkmark & \checkmark & 94.86 \\ \hline
16 &  & \checkmark &  &  & \checkmark & \checkmark &  & 93.42 \\ \hline
17 &  &  & \checkmark & \checkmark &  &  & \checkmark & 96.19 \\ \hline
18 &  &  & \checkmark & \checkmark &  &  &  & 95.59 \\ \hline
19 &  &  & \checkmark & \checkmark &  & \checkmark & \checkmark & 95.66 \\ \hline
20 &  &  & \checkmark & \checkmark &  & \checkmark &  & 96.08 \\ \hline
21 &  &  & \checkmark &  & \checkmark &  & \checkmark & 94.30 \\ \hline
22 &  &  & \checkmark &  & \checkmark &  &  & 92.25 \\ \hline
23 &  &  & \checkmark &  & \checkmark & \checkmark & \checkmark & 95.24 \\ \hline
24 &  &  & \checkmark &  & \checkmark & \checkmark &  & 89.08 \\ \hline
\end{tabular}%
}
\end{table}

Strategy 2, used during the NAS process, achieves the highest accuracy (96.59\%) and represents a frequently adopted approach in prior works. It anonymizes IP addresses without zeroing them and avoids UDP padding, retaining key features for classification. The model also performs well under more restrictive preprocessing, like strategy 6 (95.38\%), which zeros IP addresses, or strategy 7 (95.12\%), which zeroes both IP addresses and ports while applying UDP padding. These results show that the proposed model generalizes well, without overfitting specific header fields like IP addresses or ports.

Some strategies cause drops in accuracy when compared to  Strategy 2. For example, Strategy 24, which zeros MAC addresses, IP addresses, and ports, leads to the lowest accuracy (89.08\%), with a drop of 7.51\%. Strategy 22, similar to Strategy 24 but without port zeroing, performs slightly better (92.25\%) but still exhibits a 4.34\% drop, suggesting that excessive zeroing disrupts critical structural patterns. However, Strategies 23 and 21 (95.24\% and 94.30\%, respectively), using UDP padding, absent in Strategies 22 and 24, partially mitigate the accuracy drops caused by extensive zeroing.

Removing the Ethernet layer entirely, as in Strategy 8 (95.00\%), outperforms zeroing MAC addresses in Strategy 24 (89.08\%). Both strategies are identical in other preprocessing steps. This complete removal prevents partial masking and preserves structural consistency, yielding better performance. We discourage anonymizing MAC addresses, as demonstrated in Strategies 9 through 16. 
For example, Strategy 10 (95.42\%) takes a moderate approach by anonymizing both MAC and IP addresses instead of zeroing them but still yields limited gains. It is 1.17\% below the best strategy, highlighting the minimal impact MAC addresses have on classification.

The proposal performs robustly across many configurations. Strategies like 7 and 23 demonstrate its adaptability, achieving a commendable balance between privacy preservation and classification performance, making it suitable for applications with privacy regulations.
These findings help practitioners strike the right balance between obfuscating sensitive fields and keeping essential header features for optimal performance.

\subsection{Impact of session length on performance} \label{sensitivity input size}

This analysis tests the impact of session length on the trade-offs between efficiency and classification performance \added{in the VPN-NonVPN task.} Header-level strategies affect accuracy, leaving hardware measures unchanged, while session length reductions impact efficiency by altering the input size. We considered the first eight preprocessing strategies in Table \ref{tab:preprocessing_strategies} because the Ethernet layer is removed, a common practice in TC. We progressively reduced the standardized session length from 784 bytes (the standard value used in most research papers) to smaller lengths: 676, 576, 484, 400, 324, 256, and 196 bytes. The reductions are selected to capture meaningful performance trends while preserving key session-level features, such as critical header fields and early-packet information essential for classification.

Table~\ref{tab:input_size_performance} shows how different input sizes, listed in the columns, affect FLOPs and maximum tensor size (M-Tens.), listed in the rows, regardless of preprocessing strategy. As input size decreases, FLOPs drop from 10.08M to 2.24M, and M-Tens. reduces proportionally from 20.12K to 4.90K. Smaller input sizes improve efficiency, making the model more suitable for deployment in resource-constrained environments. The number of parameters remains constant at 88.26K.

\begin{table}[htbp]
\centering
\caption{Impact of input size on hardware metrics.}
\label{tab:input_size_performance}
\renewcommand{\arraystretch}{1.2}
\large
\resizebox{\columnwidth}{!}{%
\begin{tabular}{
>{\centering\arraybackslash}c|
>{\centering\arraybackslash}c|
>{\centering\arraybackslash}c|
>{\centering\arraybackslash}c|
>{\centering\arraybackslash}c|
>{\centering\arraybackslash}c|
>{\centering\arraybackslash}c|
>{\centering\arraybackslash}c|
>{\centering\arraybackslash}c
}
\hline
\textbf{Input Size (B)} & \textbf{784} & \textbf{676} & \textbf{576} & \textbf{484} & \textbf{400} & \textbf{324} & \textbf{256} & \textbf{196} \\
\hline
\textbf{FLOPs (M)} & 10.08 & 8.61 & 7.25 & 6.07 & 4.90 & 3.95 & 3.01 & 2.24 \\
\textbf{M-Tens. (K)} & 20.12 & 17.29 & 14.71 & 12.38 & 10.19 & 8.26 & 6.45 & 4.90 \\
\hline
\end{tabular}%
}
\end{table}

\begin{figure}[htbp]
    \centering
    \includegraphics[width=\columnwidth]{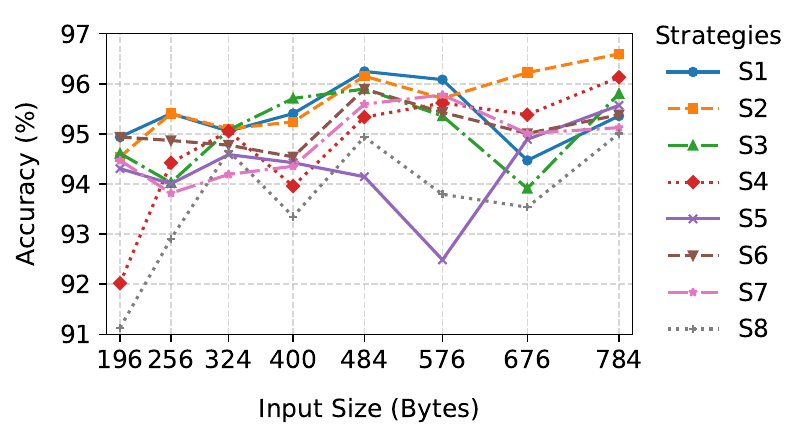}
    \caption{Accuracy across strategies for different input sizes.}
    \label{fig:input_size_variation}
\end{figure}

Figure \ref{fig:input_size_variation} complements Table \ref{tab:input_size_performance} by showing how accuracy varies with input size across the first eight strategies from Table \ref{tab:preprocessing_strategies}. The x-axis represents input size, while the y-axis shows accuracy for each strategy. The figure highlights that for most strategies (1, 2, 3, 5, 6, 7) a small drop of 1 to 2\% at 196 bytes affected the accuracy. For strategies with extensive zeroing, such as Strategy 8 (zeros IP addresses and ports) and Strategy 4 (zeros ports but anonymizes IP addresses), the accuracy drop becomes more pronounced, reaching up to 4\% as input size decreases. Conversely, Strategies 7 and 3, which follow the same steps as 8 and 4, respectively, but include UDP padding, mitigate this issue. Incorporating padding mechanisms can trade off input size reduction and classification performance, as shown by the table and figure.

\section{Embedded Inference Deployment} \label{deployment}

\subsection{\protect\added{Deployment Methodology and Setup}}

\added{Building on the session length analysis in Section~\ref{sensitivity input size}, we evaluate how the model selected by HW-NAS performs on MCUs across various input sizes. All deployment results use this architecture and preprocessing Strategy~2 from Table~\ref{tab:preprocessing_strategies}, which served as the main configuration throughout this study.}

\protect\added{We consider a deployment setting in which fixed-length session vectors are prepared on an upstream device (e.g., an IoT gateway or embedded Linux board), 
\superadded{which is assumed to perform packet capture, session aggregation, and buffering}, 
and forwarded to a microcontroller (e.g., STM32F401RE or STM32F746G), 
\superadded{which executes the classification stage under tight memory and compute constraints}. 
\superdeleted{for lightweight inference.}\superadded{Inference is triggered when the required input representation is available, and real-time operation is therefore defined in terms of bounded response time at the session level.}
Our evaluation focuses on MCU-side execution, which represents the actual resource-constrained component of the deployment pipeline \superadded{and determines the feasibility of on-device TC}.}

\added{The HW-NAS-optimized architecture was quantized to 8-bit (INT8) using TensorFlow Lite and converted to C code via STM32Cube.AI. 
\superadded{Both post-training quantization (PTQ) and quantization-aware training (QAT) were considered; PTQ applies INT8 conversion to a pretrained Float32 model, while QAT fine-tunes the pretrained model under simulated INT8 arithmetic to mitigate quantization-induced accuracy loss.} \added{For each input size, \superadded{PTQ and QAT were} repeated 10 times to evaluate accuracy variability; one representative INT8 model per input size was deployed, as all quantized variants share the same architecture, memory footprint, and computational cost.}
}

\added{Inference metrics were averaged over 1000 runs per device. A USB power meter, connected between the board and host PC, recorded current and voltage during idle and active phases. Power per session was derived from the average current and voltage during the active phase, and energy was calculated as $E = P \cdot t$, where $t$ is the inference latency.}

\subsection{\protect\added{Experimental Results and Discussion}}

\begin{figure}[htbp]
  \centering
  \includegraphics[width=0.95\columnwidth]{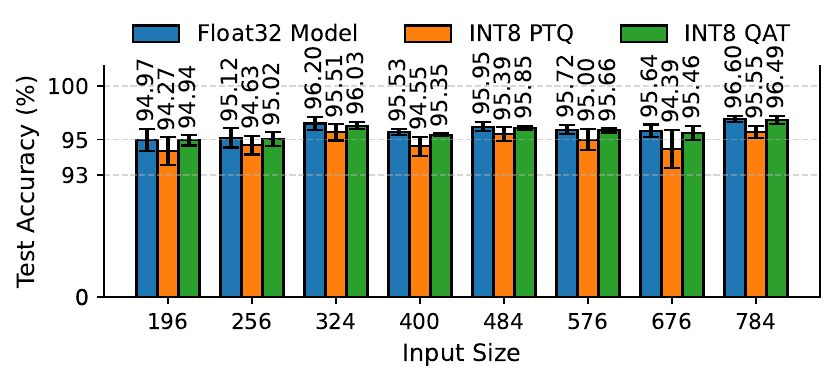}
  \caption{\protect\superadded{Accuracy across quantization levels (VPN-NonVPN).}}

  \label{fig:quantization_accuracy}
\end{figure}

\protect\added{Figure~\ref{fig:quantization_accuracy} compares model accuracy across Float32 and INT8 configurations \protect\superadded{(PTQ and QAT)} for input sizes from 196 to 784 bytes. 
The x-axis reports the input size, while the y-axis shows test accuracy (\%). 
Error bars indicate the standard deviation across 10 independent training and quantization runs.}

\superdeleted{Post-quantization accuracy remains high, with absolute drops under 1.25\%. The small, consistent degradation confirms stable performance suitable for deployment on resource-constrained embedded devices.}

\superadded{INT8-PTQ incurs a moderate but consistent accuracy degradation; QAT recovers most of this loss and significantly reduces variance across runs. 
Across all input sizes, QAT remains within 0.1--0.3\% of the Float32 baseline, confirming stable and deployment-ready behavior.}

\begin{figure}[htbp]
  \centering
  \includegraphics[width=0.95\columnwidth]{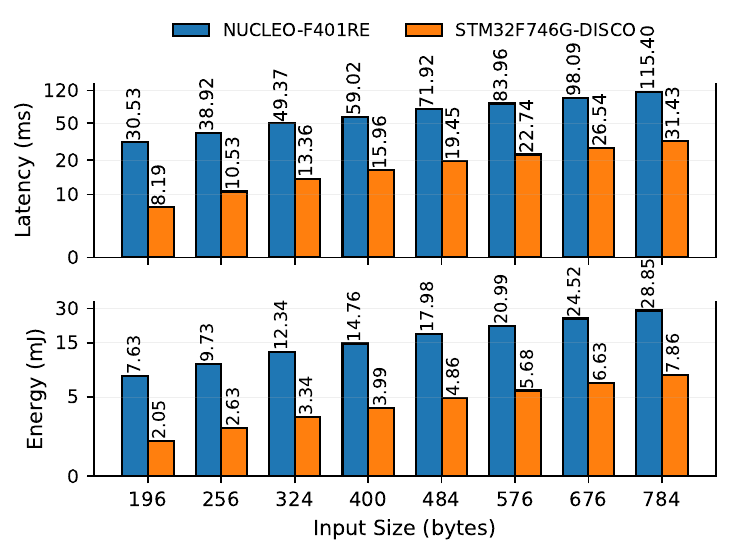}
\caption{\protect\added{Inference cost across MCUs and input sizes.}}
  \label{fig:quantization_latency_energy}
\end{figure}

\added{Figure~\ref{fig:quantization_latency_energy} illustrates inference latency (top, in ms) and energy per session (bottom, in mJ) across input sizes and MCUs.}

\added{During inference, both boards exhibited an average current draw of $0.05$\,A at a 5\,V supply, resulting in an average active power of $0.25$\,W across the tested configurations. Idle current was measured separately: $0.02$\,A for the F401RE and $0.23$\,A for the F746G, corresponding to idle power baselines of $0.10$\,W and $1.15$\,W, respectively. Energy values in Figure~\ref{fig:quantization_latency_energy} are based on active power only, isolating the model execution cost from background consumption. }

\added{Latency increases with input size, as expected. The Nucleo-F401RE maintains latency below $60$\,ms for sizes up to $400$ bytes, but reaches $115$\,ms at 784 bytes. The STM32F746G achieves consistently lower latency, ranging from $8$\,ms to $31$\,ms, reflecting its faster clock and memory bandwidth. Energy trends follow latency linearly due to constant power.}

\added{At an input size of 784 bytes, the model runs at its full hardware configuration found by HW-NAS: 88.26K parameters, 20.12K max tensor, and 10.08M operations (the INT8-equivalent of the baseline FLOPs), matching the setup used throughout Section~\ref{results}. This setting yields 115.4\,ms latency and 28.85\,mJ energy on Cortex-M4, and 31.43\,ms and 7.86\,mJ on Cortex-M7. These values can serve as an empirical deployment reference for models with similar or higher complexity, assuming deployment on comparable devices.}

\added{Overall, the results confirm that HW-NAS models can be deployed on low-power MCUs for real-time TC. Input sizes like 324 and 484 bytes offer a good trade-off: they achieve over 95\% INT8 accuracy while keeping latency under 72\,ms on Cortex-M4 and 20\,ms on Cortex-M7; energy remains below 18\,mJ and 5\,mJ, respectively. Smaller inputs such as 256 bytes are more efficient but slightly less accurate. Despite the increase in latency with input size, delays remain acceptable in realistic scenarios, especially given that session aggregation and preprocessing occur on the gateway. These values align with response time ranges in embedded monitoring tasks~\cite{jiang2018low_flops_latency}.}

\added{
The literature includes only a few works that report deployment measurements, and the target platform varies across studies.
\superadded{Table~\ref{tab:latency_comparison} summarizes inference latency reported in prior work. Each row corresponds to a method and platform, while the columns report the hardware class, power budget (in watts), input type, and inference latency. For our work, two session input sizes are reported per platform, corresponding to a maximum-length configuration (784~B) and a shorter operating point (484~B), where B denotes bytes. Underlining highlights the best latency among low-power deployments (power budget $<1$~W), which is the primary target scenario of this work, while the asterisk ($^{*}$) denotes the best overall latency across all platforms.}
}
\begin{table}[htbp]
\centering
\caption{\superadded{Inference latency by platform and power budget.}}
\label{tab:latency_comparison}
\renewcommand{\arraystretch}{1.2}
\Large
\resizebox{\columnwidth}{!}{%
\begin{tabular}{
>{\centering\arraybackslash}c|
>{\centering\arraybackslash}c|
>{\centering\arraybackslash}c|
>{\centering\arraybackslash}c|
>{\centering\arraybackslash}c|
>{\centering\arraybackslash}c
}

\hline
\textbf{Work} &
\textbf{Platform} &
\makecell{\textbf{Hardware} \\ \textbf{Class}} &
\makecell{\textbf{Power} \\ \textbf{Budget (W)}} &
\textbf{Input} &
\textbf{Latency} \\
\hline

% ---- Our work ----
\multirow[c]{4}{*}{\textbf{Our work}} &
\multirow[c]{2}{*}{STM32F746G} &
\multirow[c]{2}{*}{Cortex-M7} &
\multirow[c]{4}{*}{$<1$} &
\multirow[c]{4}{*}{Session} &
31.43 ms {\large (784 B)} \\
 &  &  &  &  &
\underline{\underline{19.45 ms}} {\large (484 B)} \\
\cline{2-3}

 & \multirow[c]{2}{*}{STM32F401RE} &
\multirow[c]{2}{*}{Cortex-M4} &
 &  &
115.40 ms {\large (784 B)} \\
 &  &  &  &  &
71.92 ms {\large (484 B)} \\
\hline

\multirow[c]{2}{*}{\cite{chehade2025energy}} &
STM32F746G & Cortex-M7 &
\multirow[c]{2}{*}{$<1$} &
\multirow[c]{2}{*}{Session} &
31.43 ms \\
 & STM32F401RE & Cortex-M4 &  &  & 115.40 ms \\
\hline

\cite{gallo2021fenxi} &
PCIe TPU &
Edge accelerator &
$\sim$30 &
Flows &
10--30 ms \\
\hline

\cite{akem2024encrypted} &
Intel Tofino &
Switch ASIC &
$\sim$20--30 &
Flows &
$<1~\mu$s\textsuperscript{*} \\
\hline
\end{tabular}%
}
\end{table}

\added{
By comparison, STM32-based inference was previously shown in~\cite{chehade2025energy}, but only for a single input size (784~bytes), without examining empirical trade-offs across different input lengths.
FENXI~\cite{gallo2021fenxi} reports 10--30\,ms inference latency using a high-power (30\,W) tensor processing unit (TPU) attached via PCI Express, under 100\,Gbps traffic with batching and model caching.
The in-switch approach of~\cite{akem2024encrypted} achieves sub-microsecond latency at line rate by executing Random Forest models directly on Intel Tofino programmable switches; however, this relies on high-end network hardware, and the feature pipeline is limited by the P4 programming model, which restricts flexibility and applicability to broader embedded settings.
\superadded{In contrast, our results show that ETC inference is feasible on microcontrollers with a power budget below 1\,W, and that practical latency can be achieved by varying the input length to balance efficiency and response time.}
}

\section{Conclusion and Future Work}

This paper introduces a DNN optimized through HW-NAS for efficient TC in resource-constrained environments. \superadded{The study shows that incorporating hardware constraints during architecture synthesis supports reproducible and deployable design choices, where accuracy is achieved under strict efficiency requirements.} The proposed model achieves up to \deleted{96.59\%}\added{96.60\%} accuracy in session-level classification while requiring up to 444 times fewer parameters and 312 times fewer FLOPs than SOTA methods. These results highlight its suitability for IoT and edge platforms, where efficiency is critical.

The optimized architecture delivers satisfactory performance across diverse tasks and settings; \added{it also obtains consistent performance in terms of cross-dataset generalization,} confirming its adaptability to different TC challenges. \deleted{The study shows that reducing input size enhances compute efficiency, enabling deployment in environments with stricter hardware constraints without sacrificing performance.}\added{The study further shows that reducing input size improves efficiency without sacrificing performance; combined with quantization, this yields notable latency and energy gains on embedded microcontrollers.} Additionally, techniques like UDP padding mitigate accuracy loss in restrictive preprocessing scenarios, offering practical solutions.

Future work will explore broader deployment scenarios, including adaptive runtime optimization and integration with real-time network operations. 
Federated learning will also be investigated to enable decentralized training with privacy preservation and improved scalability. 
We further plan to extend HW-NAS to other traffic analysis tasks, such as intrusion and anomaly detection, to assess its adaptability across domains, 
\superadded{and to incorporate additional deployment-aware objectives, such as quantization-aware constraints and robustness considerations (e.g., adversarial perturbations and distribution shift), thereby improving reliability under practical operating conditions.} 
These directions aim to advance hardware-efficient ML for next-generation network systems.

%\section*{Acknowledgment}

% Print Bibliography
\bibliography{references_adel}

\end{document}